\documentclass{emulateapj-rtx4}
\usepackage{graphicx, amsmath}
\usepackage{bm, color}
\usepackage{booktabs}
\usepackage{multirow}

\definecolor{brown}{rgb}{0.42,0.24,0.07}
\definecolor{darkgreen}{rgb}{0.0,0.6,0.00}
 \definecolor{purple}{rgb}{0.7,0.0,0.7}

\newcommand{\ov}[1]{\overline{ #1}}
\newcommand{\op}{\:\!}
\newcommand{\tp}{\;\!}

\newcommand{\Sec}[1]{Section~\ref{#1}}
\newcommand{\Fig}[1]{Figure~\ref{#1}}
\newcommand{\EQ}{\begin{equation}}
\newcommand{\EN}{\end{equation}}
\def\St{\mbox{\rm St}}
\newcommand{\erf}{{\rm erf}}
\newcommand{\Tab}[1]{Table~\ref{#1}}
\newcommand{\vv}{\mbox{\boldmath $v$} {}}
\newcommand{\uu}{\mbox{\boldmath $u$} {}}
\newcommand{\zzz}{\hat{\mbox{\boldmath $z$}} {}}
\newcommand{\Eqs}[2]{Equations~(\ref{#1}) and~(\ref{#2})}
\newcommand{\DDD}{{\cal D} {}}
\newcommand{\Eq}[1]{Equation~(\ref{#1})}
\newcommand{\nab}{\mbox{\boldmath $\nabla$} {}}

\begin{document}

\title{Protoplanetary dust porosity and FU Orionis Outbursts: Solving the mystery of Earth's missing volatiles}
\author{Alexander Hubbard\altaffilmark{1}, Denton S. Ebel\altaffilmark{2}}
\altaffiltext{1}{Department of Astrophysics, American Museum of Natural History, New York, NY 10024-5192, USA}
\altaffiltext{2}{Department of Earth and Planetary Sciences, American Museum of Natural History, New York, NY 10024-5192, USA}
\email{ahubbard@amnh.org,  debel@amnh.org}

\begin{abstract}
The Earth is known to be depleted in volatile lithophile elements in a fashion that defies easy explanation.
We resolve this anomaly with a model that combines
the porosity of collisionally grown dust grains in protoplanetary disks with heating from
FU Orionis events
 that dramatically raise protoplanetary disk temperatures.
The heating from an FU Orionis event alters the aerodynamical properties of the dust while evaporating the volatiles. 
This causes the dust to settle, abandoning those volatiles.
The success of this model in explaining the elemental
composition of the Earth is a strong argument in favor of highly porous collisionally grown dust grains
in protoplanetary disks outside our Solar System.  Further, it demonstrates how thermal (or condensation based)
alterations of dust
porosity, and hence aerodynamics, can be a strong factor in planet formation, leading to the onset of
rapid gravitational instabilities in the dust
disk and the subsequent collapse that forms planetesimals.
\end{abstract}


\section{Introduction}

Of all meteorite classes, the elemental abundances of CI chondrites most closely match the Solar photosphere,
and those meteorites are believed to comprise the most primitive solid material found in our Solar System \citep{Lodders03}.
For that reason, these meteorites are held as the standard for
the elemental and
chemical composition of solids that condensed out of the Solar nebula, before thermal, chemical, or physical
processing took their toll.
However, the bulk silicate Earth has a significantly smaller mass-fraction of volatile
lithophile elements (low condensation temperature and
trapped in the mantle, hence measurable)
 than do CI chondrites: the relative abundance of lithophile elements with condensation temperatures
below $1400$\op K decreases with decreasing condensation temperature \citep{Palme00,McDonough03}.
 That abundance trend is inverted in Solar abundances which show enhancement in those same volatiles relative to solar twins \citep{Mel09}.

Assuming that the Earth formed out of solids which at first had Solar composition, this depletion of the volatile elements
has been a long outstanding problem in the formation of the Earth, especially since \cite{Humayun95} showed
that this decrease in volatiles does not come with an isotopic signature in potassium.
Existing theories for this volatile depletion are unsatisfactory, and explaining it is the goal of this paper.

Our model combines many disparate aspects of astrophysics and planetary sciences, from FU Orionis style outbursts and dust dynamics,
to isotope ratios and the Goldschmidt classification.  In light of the breadth of reader backgrounds, 
we have attempted to provide adequate introductions to the
different moving parts before using them in detail.
Accordingly, we include an extended introduction to the problem in \Sec{Prob-overview}, detailing the constraints that
any theory for the Earth's volatile depletion must match.  We provide a narrative description of our model in
\Sec{model_overview} to provide context for the introduction to the various sub-processes found in \Sec{particulars}.
\Sec{sec_actual} finally combines the individual pieces into a coherent whole, and \Sec{extension} extends
the model to other bodies in the Solar System.  We conclude in \Sec{conclusions}.

\section{Overview of Earth's volatile depletion} 
\label{Prob-overview}

\subsection{Laboratory constraints}

The lack of an isotopic signature in the depleted potassium implies that the depletion of the volatiles cannot have been trivially generated through partial evaporation, because that preferentially
evaporates the lighter isotopes.
A brief heating episode could result in completely depleted surface layers surrounding untouched cores if
the diffusion timescale is much longer than the heating episode, which would avoid an isotopic signature.  However,
the depletion of Earth's potassium is of the order of $80\%$, so the fully depleted surface
layer would need to have a thickness of $40\%$ of the grain's radius.  Clearly, the potassium must have diffused significantly
through the solid grains.
This means that the depletion of potassium cannot be explained through models invoking partial evaporation,
or full evaporation followed by partial condensation, which instead result in an enrichment in the lighter
isotopes \citep{Richter04}.

 This depletion also
cannot be explained by planet formation at a single temperature, as that would instead result in abundances which
are undepleted for elements with condensation temperatures above that formation temperature, and nearly uniformly
depleted for elements with condensation temperatures below that formation temperature.
If planets fed from planetesimals that formed
in regions of different temperature (and therefore had different depletion patterns), then the planets' final
depletion pattern could be  a smooth function of temperature. 
However, the Earth's volatile depletion has not been successfully reproduced by recent attempts to combine
models of protoplanetary disk chemistry with
models of planet formation \citep{Bond10,Elser12}.
 This is not surprising because
the condensation temperature range for which the depletion is noted is about $650$\op K to $1350$\op K.  Theoretical
models for disks predict temperature distributions close to $T \propto R^{-1/2}$, so the needed temperature
range would
require the proto-Earth to have fed from an annulus whose outer edge was about four times farther out than its inner edge,
double the ratio of Mars' orbit to Venus'.

A further possible cause for the Earth's volatile abundances is a late veneer provided by bombardment, occurring
 after mantle/core differentiation, which could strongly alter mantle volatile abundances.  However,
 a late veneer would supply lithophile and siderophile elements alike,
 so the temperature dependence of the abundances of volatile siderophile elements would match that
 of the volatile lithophiles.  Instead, the abundances of volatile highly siderophile elements in the Earth's mantle, normalized
to Mg in CI chondrites, are low and temperature-independent, appropriate for them having been largely trapped in the iron
core during differentiation.
This rules out a late veneer as a solution to the volatile abundance problem
\citep{Wood10, Morbidelli12}.

\subsection{Implications for Earth's formation}

The only way then to deplete the pre-Earth solids of a given volatile element is to
partially separate the pre-Earth dust and
the complementary gas at temperatures above that volatile's condensation temperature, when most of that volatile's mass is in the gas phase.  Without the separation, the gas and dust combine to match Solar composition and, after cooling, the volatiles
will recondense without any depletion having occurred.
This partial segregation must occur after isotopic equilibration between gas and dust. 
The heating and segregation are then followed by cooling the dust in its new,
 volatile depleted environment, allowing recondensation of the residual gas while leaving the dust still partially depleted.

As long as the recondensation occurs rapidly compared with the cooling time of the gas, the different isotopes
will fully recondense before a significant change in the segregation occurs, and no isotopic signature will be left.
This is important because the degree of depletion found depends on the condensation temperature, so
the level of segregation itself 
must be temperature dependent: different elements with different condensation temperatures must condense
after different degrees of segregation.

\subsection{Existing models}

\cite{Albarede09} tried to explain the Earth's volatile abundances by adapting the model of
\cite{Cassen96,Cassen01}, originally intended to
explain the volatile depletion of carbonaceous chondrites.  In this model,
the proto-Solar nebula dispersed while the disk midplane was still hot ($\sim 1000$\op K), leaving the solid material behind.
The midplane temperature was sustained through the energy released by the accretion of the gas onto the proto-Sun, 
so as the disk dispersed, the accretion flow slowed and the disk cooled.
The volatiles condensed in turn from highest condensation temperature to the lowest, while
the disk gas to dust mass ratio dropped.  In this model, the dispersal of the gas disk provided the spatial dust/gas segregation.
The degree of volatile depletion associated with the dust/gas segregation was explicitly correlated with the fraction
of gas remaining, which in turn controlled the accretion flow onto the central star and hence the gas temperature.
Thus, volatile abundances were correlated with their condensation temperatures.
\cite{Ciesla08} reviews this model, and finds
that it requires parameters that push the bounds of plausibility.

The temperature history needed for Albar\`ede's model requires that gas accreted onto the proto-Sun at
rates which do not fit well with more recent observational
constraints \citep{Hartmann96,Natta06}.  Those observations indicate relatively low accretion rates during quiescent phases
and extremely high accretion rates when protostellar systems are in outbursts.  Further, the model requires that much of the dust not move significantly radially.  However, small rocks and even boulders are somewhat tied to the accretion flow, which must be significant
to sustain the temperatures required, and experience headwind-induced inspiral \citep{Weidenschilling77}.  That means, at a minimum, that most of the dust must be locked up in very large boulders by the time the temperature
dips below $1000$\op K.  Boulders large enough to resist headwind-induced infall are also insensitive to turbulent stirring, and
so have low velocity dispersions.

Low velocity dispersions naturally lead to gravitational
instabilities in the dust disk which turn the boulders into planetesimals.
With most of the potentially solid material in the form
of large boulders or planetesimals tightly settled to the midplane, volatiles in the gas phase
will tend to condense as very small dust grains which are very well coupled to the gas.
While boulders or planetesimals are individually large, the total surface area of large dust grains in this scenario is very low, and the large grains are restricted to the midplane.  
This means that they can only encounter a small fraction of the tiny volatile-rich grains, and those tiny grains would instead
be entrained by the accretion flow onto the proto-Sun.
Thus, the volatile abundances would show a dramatic drop
for those volatiles that condensed after planetesimals or boulders formed, which is not observed.

\subsection{Rotating and speeding up the model}

The model of \cite{Albarede09} seems unlikely, in a large part because segregating the gas and dust through
radial motions is slow: the velocities are low, and the lengthscales large.
Nonetheless, the segregation of dust from gas is
 quite possible in a protoplanetary disk, even though dust grains are expected to be well coupled
to the gas.  The dust grains feel the same forces as the gas with the crucial exception of the pressure forces.
 As a result, the dust can drift relative to the gas both radially (because there is a radial gas pressure gradient)
 and vertically (due to the vertical stratification of the gas).  Further options for moving
the dust through the gas include interactions
with the stellar radiation field, such as radiation pressure or photophoresis \citep{Krauss05}.

Radiative effects, however, can only process
an optically thin dust layer, which is an inadequate mass reservoir to cause the large depletions observed in the bulk dust population
as represented by Earth.
Radial segregation draws on the radial pressure gradients, which are much weaker than the vertical pressure gradients
although this is the basis of the model of  \cite{Wasson74} as elaborated on by \cite{Cassen96}.  Any grains
large enough for radial segregation to be significant will also be well settled vertically, and subject to at least one of the streaming instability
\citep{Johansen07}, headwind induced inspiral, or direct gravitational instability.  Inspiralling grains, and their depletion, would have been lost
to the proto-Sun, while the streaming instability and gravitational instability produce planetesimals too large see significant
subsequent compositional changes.

Vertical dust/gas separation is then the remaining candidate for spatial segregation.  By invoking the much stronger vertical
pressure gradients and much smaller vertical distances, rotating the problem bypasses many of the problems faced
by radial segregation.

\section{Overview of our model}
\label{model_overview}

\begin{figure*}[t!]\begin{center}
\includegraphics[width=0.9\linewidth]{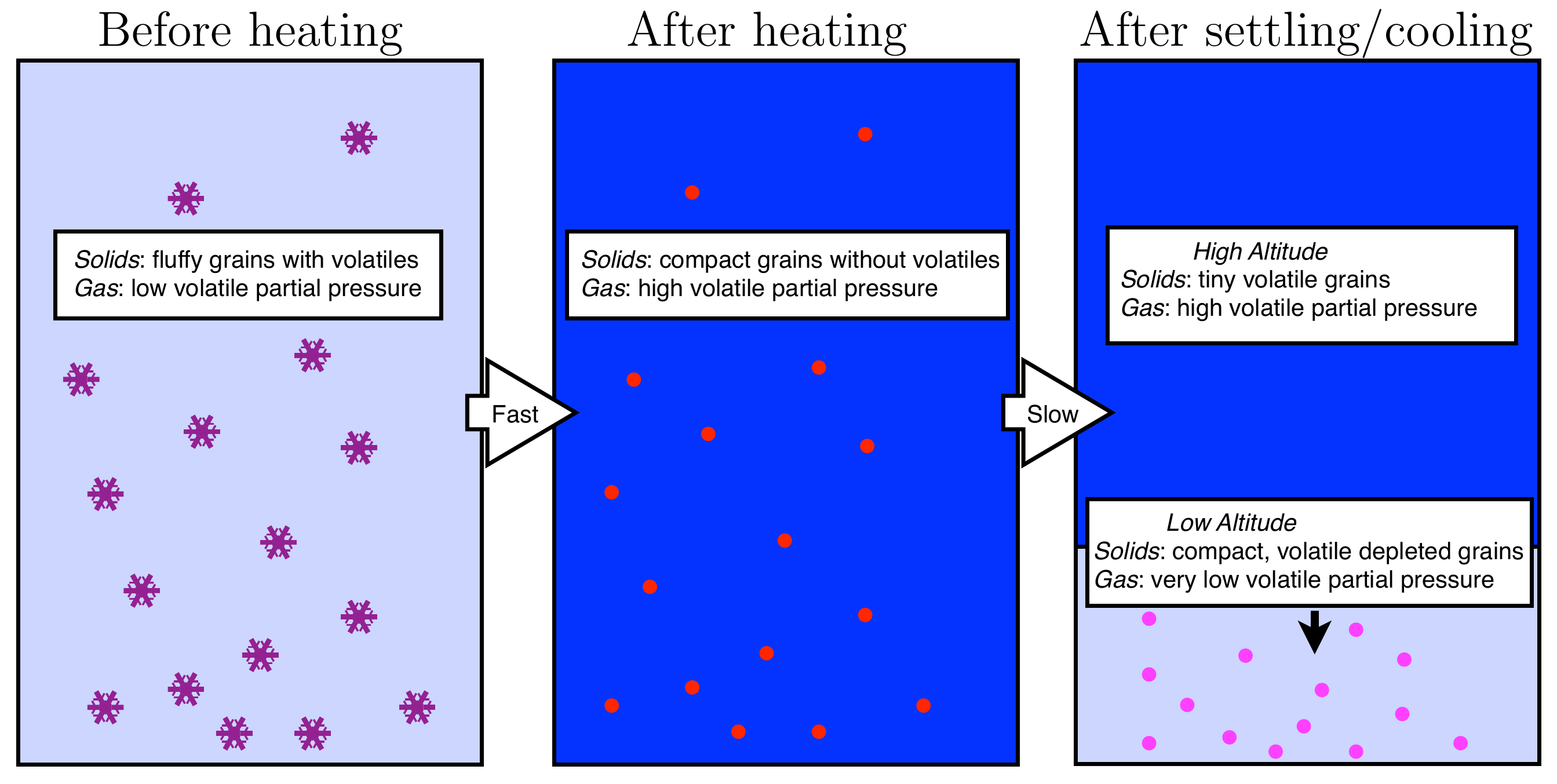}
\end{center}\caption{
Schematic of the model.
Left panel: initial state, with fluffy, vertically mixed dust in equilibrium with low amounts of volatiles in the gas.
Middle panel: immediately after heating, the dust contracts, and the volatiles evaporate, entering the gas phase.
Right panel: after settling and cooling, only the low altitude volatiles recondense into the settled grains, leaving them
depleted compared to their initial state.
\label{cartoon} }
\end{figure*}

This paper describes a process for generating vertical dust/gas segregation in a manner that is naturally temperature dependent, over
a temperature range appropriate for explaining the Earth's volatile depletion while simultaneously providing enough
time for dust/gas isotopic equilibration.  The temperature dependence of the depletion
implies that the temperature of the disk must vary strongly (at least from the temperatures of $1350$\op K to $650$\op K,
as witnessed by the condensation temperatures) on dust settling timescales.  Those settling timescales are 
themselves hard to shorten below a few tens of years, or lengthen beyond a few hundred, so the model requires
temperature fluctuations in the disk that last many orbits, but are far, far shorter than the disk lifetime.

\subsection{FU Orionis type outbursts}

Perhaps surprisingly, temperature variations of that magnitude and duration are not only possible but expected, and must
be accounted for in models of the formation of the Solar system:  FU Orionis type events (hereafter FUors) occur multiple times during protoplanetary disk lifetimes \citep{Hartmann96, Dunham08, Evans09}.  These events, which last
about a century, are caused by a large amount
of disk material being rapidly dumped onto the protostar which leads to a dramatic, four to six magnitudes, rise in
a proto-stellar system's luminosity.  Dust grains of mm to cm sizes settle vertically to the midplane
 on timescales comparable to the cooling time of FUors, suggesting a way to create temperature dependent spatial dust/gas segregation.

However, while dust settling is fast, it must compete with vertical turbulent stirring.
Increasing the vertical dust/gas segregation on short timescales therefore requires rapidly reducing
 the strength of the dust-turbulence interactions.  This implies a rapid decrease in the dust drag coefficient caused by decreasing the dust porosity: 
high porosity dust feels stronger drag and is
more strongly vertically mixed than equal mass low porosity dust.

Our model assumes that an FUor event provides the heating required to evaporate the volatiles from the dust.  For simplicity, we assume
that the region of interest for the pre-Earth solids ($R=1$\op AU) is outside the FUor engine, which means that we can
neglect the strong local accretion flow which powers the FUor and would entrain solid material. 
This also means that irradiation from the central
engine will dominate the energy budget.  This fits with a disk whose MRI-dead zone extends inwards of $1$\op AU
during quiescent phases.

\subsection{Model narrative}

The picture we propose is this (sketched in \Fig{cartoon}): at Earth's orbital position, $R=1$\op AU,
the accretion flow through the protoplanetary disk is slow and the background disk temperature is cold (the temperature
 $T \sim 280$\op K of a Minimum Mass Solar
Nebula or MMSN, \citealt{Hayashi85}).
In the disk, dust growth occurs collisionally, 
resulting in high porosity, fluffy grains which couple strongly to the turbulence and
 which are therefore well mixed vertically through the disk: the porosity can be likened to a deployed parachute,
 with low density and lots of surface area.  The porosity that naturally results from collisional grain
 agglomeration has not yet been firmly established, but very high porosities have been predicted \citep{Ormel07,Dominik07}. 
As dust grains grow, their relative (collisional) velocities are set by ever larger scale, faster turbulent motions.
Therefore, the dust collisional velocities increase as the dust grains grow, until a limiting size is reached where further
 collisions result in bouncing or fragmentation, rather than sticking and growth \citep{Zsom2010}.

Eventually, an FUor event occurs with an extremely fast accretion flow somewhere inside of $R=1$\op AU,
heating the disk and causing the temperature to spike to about $1350$\op K at $R=1$\op AU, subsequently cooling over a
single-century scale
time.  The cooling rate and the level of heating in the model are constrained by the observed volatile abundances, but are appropriate for an irradiated disk whose heating source brightened by six magnitudes, reasonable for an FUor event \citep{Hartmann96}.

This heating causes the high porosity, fluffy dust to evaporate and recondense, melt, or sinter, contracting into non-porous spheres with much lower drag coefficients:
the parachutes are stowed.
\cite{Poppe03}
found that at the temperature we expect at the Earth's position, sub-micron SiO$_2$ grains will very strongly sinter within a year:
necks between spherical grains will grow beyond the initial grain size.  While the first stages of surface sintering
do not shorten chains of stuck-together grains, and hence do not lead to compaction, sintering acts to reduce the total surface
area; and so strong sintering will eventually tighten the chains of stuck-together grains and compactify the dust \citep{Poppe03}.

At the same time as it causes the grains to contract, 
the heating also evaporates nearly all of the elements more volatile than silicon, whose $50\%$ condensation temperature is $\sim 1300$\op K \citep{Lodders03}.
Because a large fraction of the dust mass will evaporate
and recondense (Si and Mg, with a $50\%$ condensation temperature of $1327$\op K, together make up over $20\%$ of the mass of CI chondrites), nearly total
densification is guaranteed even without melting or complete sintering if grains are heated above $1350$\op K.  In our model
this occurs for grains at or inwards of $1$\op AU.  Lesser degrees of densification are expected farther out from the proto-Sun, where
the peak FUor temperature is lower.

Estimates for the diffusion coefficients of
elements through the solid grains vary, but a range of $10^{-14}$ to $10^{-9}$\op cm$^2$\op s$^{-1}$ \tp
 at the $50\%$ condensation temperature
is reasonable  \citep{Freer81, Latourette98}.  At the high end of those diffusion rates,
the cores of mm-sized non-porous grains can reach equilibrium with the gas phase over multi-year timescales.
At the low end, only micron-sized structures will experience significant diffusion, which poses little difficulty 
for evaporation as long as the grains are fractal composites made
of micron-sized grains.  Regardless, diffusion rates scale rapidly with temperature, so in our model diffusion
only limits the evaporation of elements with condensation temperatures near the peak temperature: Mg and Si.

The FUor then fades and the disk cools at the same time as the now lower-porosity dust settles, leaving behind volatiles in the upper regions of the disk,
which become dust-depleted.  This means that when the temperature drops low enough for a
given volatile element to recondense, the abundance of that volatile recondensed on the dust grains
is depleted by the amount left behind at higher altitude. 
While the volatiles at high altitude will eventually
recondense as well, they will do so slowly, creating only small dust grains because the bulk of the potential solid
material is trapped in the large, well settled grains.  These small, volatile rich grains, extremely well coupled to the
gas, remain trapped at high altitudes for long periods and are eventually lost with the gas or as a result of 
radiative processes such as photophoresis once the gas density drops \citep{Krauss05}.

The large dust grains will become subject to the Streaming Instability \citep[SI,][]{Johansen07} because
they both have a long stopping time ($\St \simeq 0.1$) and, once settled, will be in a region with
a high dust-to-gas mass ratio (at least $0.14$, or $25 \times$CI in the framework of \citealt{Ebel00}).
The streaming instability, which operates on a timescale of tens of orbits,  in turn further concentrates the dust grains, triggering gravitational
instabilities and fully locking
in the volatile depletion through the rapid creation of planetesimals.  The planetesimals are too big to meaningfully accrete
the tiny residual recondensed volatile grains because the planetesimals have a low total surface area, are highly settled and
are large enough that the gas flows around them, deflecting tiny grains \citep{Lambrechts12}.
This final step is vital.  The gas must eventually cool, and radial motion is relatively slow compared to vertical mixing.  It follows
that the recondensed volatiles will find a way back into the volatile-depleted solids if no mechanism exists to prevent them from
doing so.
 
\subsection{Extensions of our model}

In addition to explaining the volatile abundances of Earth, this mechanism also provides a
route to early planetesimals and rapid planet formation.  The same porosity reduction--settling--streaming instability mechanism
also applies to ices at and beyond Saturn's orbit, leading there not to volatile depletion, but only to
rapid planetesimal formation.  Indeed, given the known existence of heating events associated with episodic accretion,
the sensitivity of dust porosity to its temperature history is a hitherto unexplored powerful player in planet formation as
long as collisional agglomeration leads to highly porous dust, as suggested by the depletion of Earth's volatiles: even in
orbital locations (Mercury, Jupiter) where the heating leads to the destruction of most or all of the dust (silicate grains
in the case of Mercury, ice grains in the case of Jupiter), the subsequent recondensation will result
in very different grain geometries than the original process of agglomeration of sub-micron interstellar grains.
This can occur with any adequate heating source, and chondrules in the meteoritical record testify that
chondrites for near regions of the Solar Nebula that had excursions above $1600$\tp K which were not due
to FUors.

\section{Model particulars}
\label{particulars}

In this section we lay out the particulars of our model.
In \Sec{gas_dens} we derive the vertical gas density profile.
In \Sec{dust-gas} we describe how the dust interacts dynamically with the gas,
and in \Sec{dust-dens} use that to derive the dust steady-state vertical distribution.
In \Sec{sub_SI} we describe the Streaming Instability, which can rapidly turn strongly
settled dust populations into planetesimals.
In \Sec{sec_turb} we discuss when and where turbulence is expected in the gas disk and
in \Sec{sub_FUor} we introduce FU Orionis type events.
Finally, in \Sec{Earth'sCompostion} we describe the Earth volatile abundances and
in \Sec{sub_evap_cond} we discuss the details of evaporation and condensation.

\subsection{Gas vertical density profile}
\label{gas_dens}

In a protoplanetary disk in cylindrical coordinates, so long as $z \ll R$,
where $z$ is the height above the midplane and $R$ is the distance from the central object,
 the vertical projection of the gravity of the central object is:
\EQ
g_z=-z \Omega_K^2, \label{grav}
\EN
where $\Omega_K \equiv (G M_{\star}/R^3)^{1/2}$ is the Keplerian frequency around a central object
of mass $M_{\star}$.  We will use Orb$\equiv 2\pi/\Omega_K$,
 the local orbital period, as our time unit.
A vertically isothermal disk at temperature $T$ and with negligible mass compared to the central object
therefore has a vertical gas density profile $\rho_g$ given by
\EQ
\rho_g(z)=\rho_g(0) e^{-{z^2}/{2H^2}},
\EN
where $H \equiv c_s/\sqrt{\gamma}\, \Omega_K$ is the gas scale height in hydrostatic equilibrium,
$\gamma$ is the adiabatic index and $c_s$ is the sound speed.
Under those conditions, the fraction of the vertically integrated gas mass contained within the column extending
between $z= \pm h$ is
\EQ
f(h/H)=\erf (2^{-1/2} h/H), \label{gasfraction}
\EN
see \Tab{facesof}.

\begin{table}[htbp]
\caption{The different faces of $f$ \label{facesof}}
\centerline{\begin{tabular}{ll}
\hline
\multirow{2}{0.15\columnwidth}{$f(h/H)$} & \multirow{2}{0.8\columnwidth}{The mass fraction of the gas disk contained in the column of height $z$ (Eq \ref{gasfraction}).} \\ \\
\hline
\multirow{3}{0.15\columnwidth}{$\ov{f}(T)$} & \multirow{3}{0.8\columnwidth}{An analytical fit to measurements of Earth's abundance of volatiles with $50\%$ condensation temperature $T$, normalized to CI chondrites (Eq \ref{Earth}). } \\ \\ \\
\hline
\multirow{6}{0.15\columnwidth}{$\widetilde{f}(t)$} & \multirow{6}{0.8\columnwidth}{The ratio of the gas mass within the (partially settled) column of dust at time $t$ to the gas mass within the (unsettled) initial dust column. In our model, this is also the abundance of volatiles which condense at time $t$ normalized to
the initial dust composition, assumed to be CI chondritic (Eq \ref{frac}).} \\ \\ \\ \\ \\ \\
\hline
\multirow{3}{0.15\columnwidth}{$f_m(\alpha,\St)$} & \multirow{3}{0.8\columnwidth}{The volatile abundance floor that results from
turbulent mixing with strength $\alpha$, given a Stokes number $\St$, due to the mixing height $h_m$ (Eq \ref{fm}).} \\ \\  \\
\hline
\end{tabular}}
\end{table}

\subsection{Dust-gas interactions}
\label{dust-gas}

Dust grains interact with the gas according to a drag equation
\EQ
\frac{\partial \vv}{\partial t} = -\frac{\vv-\uu}{\tau} -g_z \zzz \label{drag}
\EN
where $\vv$ is the dust grain's velocity and $\uu$ is the gas velocity at the dust grain's position.  The stopping time $\tau$
is determined by the drag regime, the mass of the dust grain, and its effective cross section with respect to the gas.  In this paper
we always assume that we are in the Epstein drag regime, i.e.~particle size small compared to the gas mean free path;
we will check this \emph{a posteriori} in \Sec{grainsize}.  In that case, we have
\EQ
\tau(z)=\left[\frac{\rho_{d,s}}{\rho_g(z)}\right]\left[\frac{a}{c_s}\right]=\tau_0 e^{\tp {z^2}/{2H^2}}, \label{tau}
\EN
where $\rho_{d,s}$ is the dust solids density (not to be confused with $\rho_d$, the dust fluid density), $a$ is the dust grain
effective radius, $c_s$ is the gas sound speed, and $\tau_0$ is the dust stopping time at the midplane.
Note that the total gas surface density
\EQ
\Sigma = 2 \pi \rho_0 H \propto \rho_0 c_s, \label{eq_Sigma}
\EN
so $\tau_0 \propto \Sigma^{-1}$.
The stopping time $\tau$ is normalized to orbital timescales through the Stokes number $\St \equiv \tau \Omega_K$.
We will write $\St_0 \equiv \Omega_K \tau_0$.

\Eqs{grav}{drag} imply that dust grains settle towards the midplane at a terminal velocity
\EQ
v_s(z,\St) = -z \Omega_K \St, \label{terminal}
\EN
which defines a local settling time 
\EQ
t_s(z,\St)=|z/v_s|=1/\St \Omega_K. \label{ts}
\EN
Because $\tau(z)$, and hence $\St$ and $t_s$, depends on the gas density which varies with the height above the midplane, 
it is convenient to also define the midplane settling time using the midplane gas density: 
\EQ
t_{s0}(\St_0)=t_s(0,\St_0)=1/\St_0 \Omega_K.
\EN
As we will see, in our models $t_{s0} \gtrsim 30$ Orb, so $\tau_0 \lesssim 10^{-3}$ Orb, and the dust is well coupled to the gas.

\subsection{Dust: vertical mixing}
\label{dust-dens}

Protoplanetary disks are generally thought to be turbulent, and turbulence mixes the dust vertically, counteracting settling.
\cite{Takeuchi02}
calculated the vertical dust density profile assuming a vertical dust turbulent diffusion coefficient $\DDD_d = \DDD_g/$Sch, where
Sch is the vertical Schmidt number and $\DDD_g$ is the gas turbulent diffusivity.  Estimates for the vertical Schmidt number
in protoplanetary disks vary, but we will normalize our Schmidt numbers to $\text{Sch}=3$ when needed \citep{Johansen06}.
The $\alpha$--disk prescription of \cite{SS73} normalizes the gas turbulent diffusivity to its pressure, so that 
\EQ
\DDD_g=\alpha c_s H,
\EN
where $\alpha$ parameterizes the strength of the turbulence.  A reasonable range for $\alpha$ in protoplanetary disks
is $10^{-4}-10^{-2}$ \citep{Blackman08}.

Takeuchi \& Lin found
\EQ
\epsilon(z)=\epsilon(0) \exp\left[-\frac{\text{Sch}\tp \St_0} {\alpha}\left(\exp\frac{z^2}{2H^2}-1\right)\right], \label{settling}
\EN
where $\epsilon$ is the local dust-to-gas mass ratio.
\Eq{settling} implies that dust is well mixed with the gas up to a mixing height
\EQ
h_m(\alpha/\text{Sch}\tp \St_0)= \left(2\ln\left[1+ \frac{\alpha}{\text{Sch}\tp\St_0}\right]\right)^{1/2} H. \label{hm}
\EN
Note that $h_m$ depends both on the strength of the turbulence through $\alpha$ and on the drag
characteristics of the dust through $\St_0$.
In what follows we will approximate \Eq{settling} by $\epsilon(z)=\epsilon(0)$ for $z\le h_m$ and $\epsilon(z)=0$ for $z > h_m$,
i.e.~assuming that all the dust is below the mixing height. 

\begin{table}[htbp]
\caption{Parameters and variables. \label{params}}
\centerline{\begin{tabular}{ll}
\hline
$h_m$ & Turbulent mixing height (Eq \ref{hm}) \\
$h_{m-}$ & Initial height of the dust column \\
$h_{\text{MRI}}$ & Top of the MRI dead zone (if applicable) \\
$\tau$ & Dust stopping time (Eq \ref{tau}) \\
$f$, $\ov{f}$, $\widetilde{f}$ & See \Tab{facesof} \\
St & $\tau \Omega_K$, the Stokes number \\
$t_s \equiv -z/v_s$ & Dust settling timescale (Eq \ref{ts}) \\
$H = c_s/\sqrt{\gamma} \Omega_K$ & Gas scale height \\
$\Sigma_g$ & Gas surface density \\
$\alpha_0$, $\alpha_1$, $\alpha_2$ & Initial, over and sub $1000$\op K turbulence $\alpha$s \\
$\rho_{d,s}$ & Dust grain material density \\
$T_0$, $T_H$, $t_c$ & Initial temp., FUor heating, FUor cooling time \\
$\epsilon$ & local dust-to-gas mass ratio \\
\hline
\multicolumn{2}{l}{Subscripts $-$ and $+$ refer to pre and post FUor values.} \\
\multicolumn{2}{l}{Subscript $0$ refers to midplane values for $\tau$, St, and $t_s$.} \\
\hline
\end{tabular}}
\end{table}

\subsection{Streaming instability}
\label{sub_SI}

Due to the radially decreasing density and temperature profiles of protoplanetary disks, the pressure is also
radially decreasing.  The resulting outwards-pointing pressure force causes the gas to orbit at a slightly
sub-Keplerian speed,
while the dust, which does not feel the pressure force, would naturally orbit
at the Keplerian speed.  This causes the dust to feel a headwind, whose effect on the dust dynamics peaks
for dust with $\St \sim 1$.
However, in the Streaming Instability (SI) introduced by \cite{Johansen07}, clumps of dust can
self-shield from the headwind.  This leads to local concentrations of dust grains far above the volume averaged
dust-to-gas mass ratio, which
can rapidly grow large enough to be gravitationally unstable, resulting in planetesimals.

This instability requires relatively large values of $\St$, and dust-to-gas mass ratios somewhat elevated above the canonical
value of $0.01$ associated with interstellar gas.  Our best fit model predicts a settled layer with a dust-to-gas mass
ratio above $0.14$ and a final $\St \simeq 0.07$, which is adequate to trigger the SI,
although by a small enough margin that small differences in the initial dust-to-gas mass ratio, often taken
to be $0.005$ in the absence of water ice, could significantly slow the SI's growth.  Using \cite{Youdin05} and \cite{Youdin07}, we estimate
 that the SI
will grow on a timescale of $10$--$30$ Orb, faster than the cooling time and comparable to the settling time.

The resulting rapid planetesimal formation is vital because it locks in the elemental
compositions predicted by our model before further turbulent mixing and diffusion can erase them: the volatiles
are still present in our model, just at a higher altitude.  Unlike dust grains, planetesimals are well beyond the Epstein drag
regime, and gas flow around the planetesimals will entrain any tiny volatile-rich dust, preventing that volatile rich dust from
rejoining the volatile depleted planetesimals \citep{Lambrechts12}.  Planetesimal formation is also important
because planetesimals are too large to interact meaningfully with the accretion flow, and do not experience
headwind-induced inspiraling.  Without this planetesimal formation, the evidence for an FUor in the dust at $R=1$\op AU will be rapidly erased as the dust is lost to the star.

\subsection{Presence of turbulence}
\label{sec_turb}

There are two main candidate drivers for turbulence in protoplanetary disks: gravitational instability (GI) and the magneto-rotational
instability (MRI, \citealt{BH91}).  The GI has long been suspected to play a role in FUor events \citep{Armitage01}, but is
only active in massive disks.  If the disk were subject to the GI, that instability would rapidly end as the accretion flow of the FUor
event dumped the mass needed for the instability onto the central star.  Similarly, at Earth's position in a protoplanetary disk
the MRI is not generally expected to be active at the midplane because the disk ionization is too low:
the column density is too high for ionizing radiation to penetrate to the midplane while
the background temperature of $T_0 \sim 280$\op K is too low for thermal ionization, which requires temperatures above
$1000$\op K \citep{Gammie96}.  Instead there are MRI active layers at the surface where non-thermal ionization allows the gas
to couple to the magnetic fields.

We assume that in the pre-FUor phase, at $R=1$\op AU, we are in the layered accretion scenario \citep{Gammie96}, with higher $\alpha$
values in the active surface layers and lower $\alpha$ values in the denser midplane.
\cite{Oishi09} found that even when the MRI dies in the midplane, the thin MRI-active layers at the surface of the disk
generate significant hydrodynamical motion throughout the disk.  The flows in those simulations were, however, largely
oscillatory, and are not expected to be particularly strong vertical mixers.
However, as long as the midplane's $\alpha$
value leads to a dust scale height $h_m> h_{\text{MRI}}$,
where $h_{\text{MRI}}$ is the base of the active layer, the vertical $\alpha$ variation will not change the dust's vertical distribution,
which then depends only on the active-layer's $\alpha$ value.
Once the FUor occurs and
the disk heats above $1000$\op K, the full disk height will become MRI-active.  

We therefore assume that the value of the viscosity parameter $\alpha$ is relatively constant before the outburst and and 
during the cooling, until the temperature
drops below $1000$\op K.  At that point the midplane ceases to be adequately ionized to support the MRI
and its turbulence rapidly decays.
We will write the initial (dating from the events leading up the FUor) strength of the active layer turbulence as $\alpha_0$,
the strength of the turbulence at the $T=1000$\op K threshold as $\alpha_1$,
and the strength of the midplane turbulence below the threshold (MRI-dead) as $\alpha_2$,
assuming $\alpha_0 \gtrsim \alpha_1 \gg \alpha_2$. 
While non-thermal irradiation will recreate a layered-accretion scenario, the dust will have settled and
compactified, and we assume that then, $h_m<h_{\text{MRI}}$ for the midplane turbulence.  This assumption
is checked in \Sec{max_turb}.

\subsection{FU Orionis type events}
\label{sub_FUor}

FUor events (see \citealt{Hartmann96} for a review) are brief, circa century length events occurring in protostellar systems
during which the accretion disk outshines the central star by factors of $100$--$1000$. Due to the rate at which they have
been observed to occur, it is estimated that they occur $5$--$10$ times per protostar \citep{Hartmann96}.  As such, any model of the formation
of the Solar System must acknowledge the near-certainty that multiple such events occurred during the Solar nebula's lifetime.
The events are thought to
be caused by an episode of greatly enhanced mass accretion through the disk
onto the protostar, up to about $10^{-4} M_{\odot}$\op yr$^{-1}$.

This enhanced mass flow heats the disk before decaying away on a timescale of many decades.
In this paper, we will assume an FUor event triggering at $t=0$. 
The time dependence of the temperature
response to such an event can be modeled as
\EQ
T(t)=T_0+T_H e^{-t/t_c}. \label{temp}
\EN
where $T_H \sim 1100$\op K is the maximal heating
from the outburst and $t_c \sim 100$\op yr is the cooling time.
The rise times for FUors are far shorter than other dynamical times for this problem, on the scale of a year \citep{Hartmann96}, 
so we neglect them.
We will use $T_0 \sim 280$\op K as the quiescent disk temperature at Earth's location \citep{Hayashi85}.
In our model, the peak temperature is constrained by the need to marginally deplete Si, but not Mg ($50\%$ condensation temperatures of
$1302$\op K and $1327$\op K respectively).

For simplicity we assume that at Earth's location the
disk is primarily heated by irradiation from the very inner disk, as this results in a vertically isothermal temperature structure
as long as the vertical thermal equilibration timescale is fast enough.
The heating used is appropriate
for an approximate six-magnitude brightening, well within the range for FUors \citep{Hartmann96}.
This assumption is not unreasonable, as \cite{Bae13} find that irradiation can be comparable to local accretion heating,
and the accretion mechanism of FUors, and its radial extent, is not well constrained.  We assume that our region of interest
lies outside the FUor engine which allows us to
neglect the accretion flow of the gas: beyond the active
FUor accretion region the local gas accretion rate is well below the $10^{-4} M_{\odot}$\op yr$^{-1}$ of an FUor.

Existing models for FUors often place the accretion region outside of $R=1$\op AU \citep{Zhu10,Bae13}.  However models for FU Orionis events
invoke small radial mismatches in accretion rates that lead to large mass pileups. 
The theoretical prediction of this small mismatch depends very strongly of the detailed behavior of protoplanetary accretion disks on long (secular) timescales.
Hence, our current limited understanding of the fundamentals of accretion in these physically complicated systems leads to a significant uncertainty in these FU Orionis models.

Irradiation can only set the temperature vertically through a disk if the thermal diffusion time is sufficiently faster than
the FUor time.  In the optically thick limit, the diffusive thermal energy flux is
\begin{equation}
{\bm F} = -\frac{16 \sigma_B T^3}{3 \rho \kappa} \nab T,
\label{radflux}
\end{equation}
where  $\sigma_B$ is the Stefan-Boltzmann constant and $\kappa$ the Rosseland mean opacity.  Combining
\Eq{radflux} with the relation $T=(\gamma-1)\bar{m} e/\rho k_B$, where $\bar{m}$ is the mean gas particle mass, we can
find an effective temperature diffusion coefficient 
\EQ
\mu \equiv \frac{16 (\gamma-1)  \bar{m} \sigma_B T^3}{3 \rho_0^2 \kappa k_B}.
 \label{therm_diff_eff}
 \EN
Note that while \Eq{therm_diff_eff} can be used to estimate temperature diffusion timescales, temperature's time dependence does not
actually obey a diffusion equation.  We can combine \Eqs{eq_Sigma}{therm_diff_eff} to estimate the temperature equilibration time
for a scale height at the midplane is
\EQ
t \simeq \frac{H^2}{\mu} = \frac{3 \kappa k_B \Sigma^2}{64 \pi^2(\gamma-1) \bar{m} \sigma_B T^3}.
\label{t_temp_diff}
\EN
Using $\Sigma=1700\tp \text{g\op cm}^{-2}$, $\bar{m}=2.33$\tp amu, $T=1000$\tp K and $\gamma=1.42$, this reduces
to
\EQ
t=0.68\tp \kappa_{\text{cgs}} ~ \text{yr}.
\EN
Irradiation can therefore control the temperature as long as $\kappa$ is at most several cm$^2$\tp s$^{-1}$.  
Opacity is dominated by tiny dust grains, and recent work has shown both that only a small fraction of dust is
found in such nebular fines \citep{Birnstiel12}, and that nebular fines can be destroyed by the temperatures we consider
\citep{Wasson08}.  This can reduce the opacity from older estimates by over two orders or magnitude
\citep{Flock13}, so $\kappa=1$\op cm$^2$\tp s$^{-1}$ is a reasonable value for our purposes.  Note that the timescale
in \Eq{t_temp_diff} scales as $\Sigma^2/T^3$, so if $\Sigma \propto R^{-1}$ and $T\propto R^{-1/2}$, the thermal vertical equilibration
time will be lower at larger $R$.

\subsection{Earth's composition}
\label{Earth'sCompostion}

In Figures \ref{Comp}, \ref{Comp2}, and \ref{Comp3} we show selected lithophile volatile abundances of the Earth relative to
those of primitive CI chondrites \citep{Lodders03,McDonough03}, as a function of $50\%$ condensation temperature at $P=10^{-4}$ bar.  Lithophile abundances are chosen because they are chemically trapped within the mantle of the Earth and their
abundances can therefore be directly determined.  CI chondrites are used as the reference because they
are the most primitive Solar System material
\citep{McDonough03}, and therefore our best record for the elemental composition of solids
condensed out of the Solar nebula \citep{Lodders03}.  All abundances are normalized to the Mg abundances, as those show little
variation across chondrite types \citep{Palme03}, which means that we explicitly assume that Mg is precisely
undepleted.  We show in \Sec{model} (Fig. \ref{Comp3}) that our model is robust to deviations from this hypothesis.
We approximate these lithophile volatile abundances of silicate Earth with respect to CI chondrites as
\EQ
 \ov{f}(T)=\min \left[1, A \left(\frac{T}{T_0}\right)^B\right]; \label{Earth}
 \EN
 see \Tab{facesof}, with $A \simeq 1.9 \times 10^{-3}$, $B \simeq 4.15$ and $T_0=280$\op K.
See \Fig{Comp} for a comparison of the fit from \Eq{Earth} to the observations.

\begin{figure}[t!]\begin{center}
\includegraphics[width=\columnwidth]{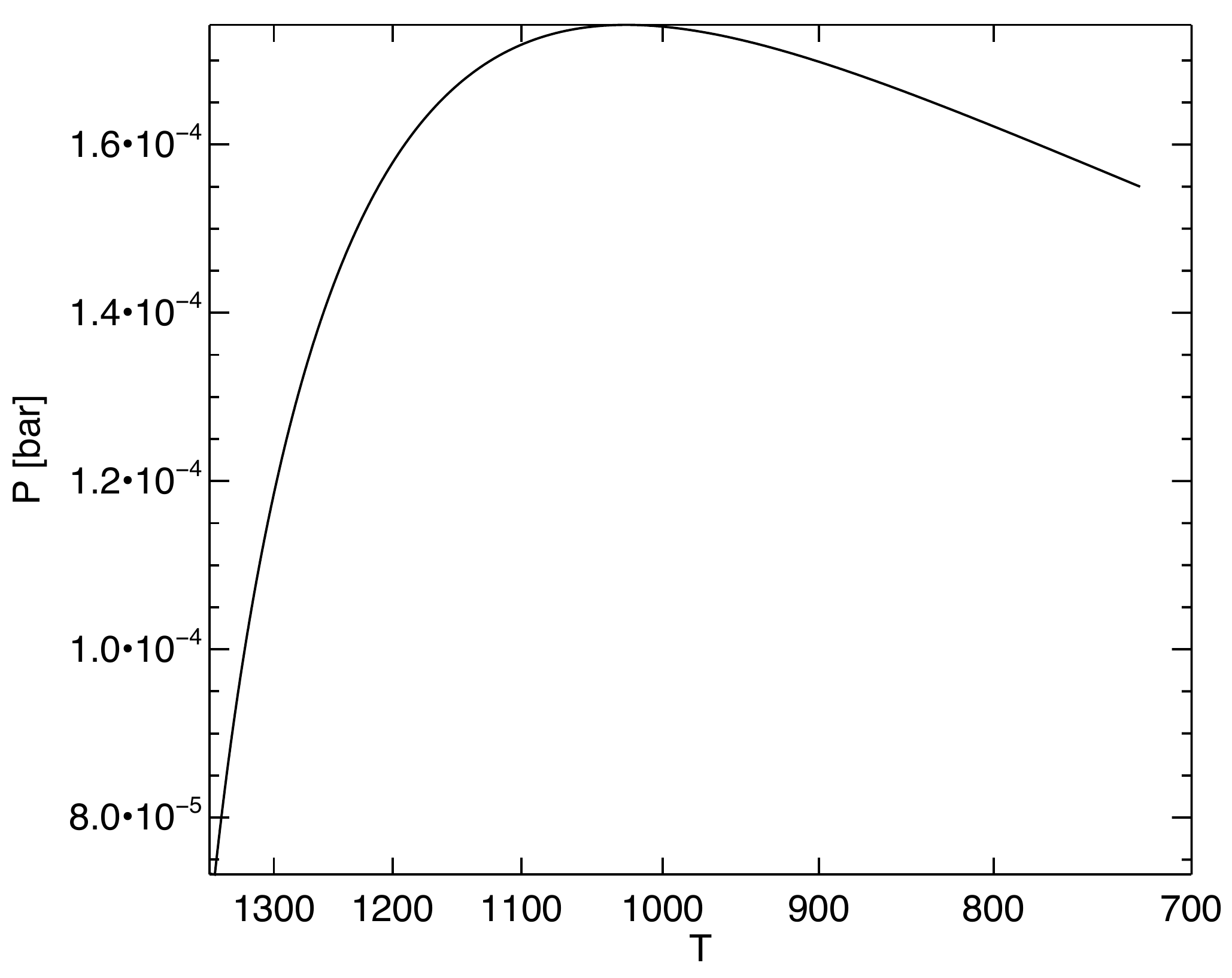}
\end{center}\caption{
The pressure experienced by the top of the dust column as a function of disk temperature (our best fit model assuming an MMSN
gas surface density).
\label{pressure} }
\end{figure}

\subsection{Volatile evaporation and condensation}
\label{sub_evap_cond}

It should be noted that the $50\%$ condensation temperatures are pressure dependent.
$P=10^{-4}$\,bar is a good estimate for the disk mid-plane pressure for a MMSN at $R=1$\op AU, at $T=280$\op K. 
Our model is at higher temperature, but lower density than that, so in \Fig{pressure} we show the pressure experienced by the top of
the dust column (which settles over time)
as a function of disk temperature for our best fit model.  This demonstrates that $P=10^{-4}$\,bar is not a bad approximation for
our purposes. 

When we extend our model to Mars, we will assume that a factor of $3$ decrease in pressure
corresponds to a decrease of $25$\op K in condensation temperature, adapted from the results of \citealt{Ebel06}, Plate 1.  Additionally, as
 we will show in \Sec{model}, 
systematic errors in the condensation temperatures can still be fit in our model by varying the parameter $T_H$, the increase
in temperature associated with the Fuor.

One assumption we need to make is that the diffusion time of the evaporated elements through
the dust grains is also short.
As noted earlier, values for the diffusion coefficients of
elements through the solid grains vary, but a range of $10^{-14}$ to $10^{-9}$\op cm$^2$\op s$^{-1}$ \tp
 at the $50\%$ condensation temperature
is reasonable  \citep{Freer81, Latourette98}, and diffusion speeds up rapidly above that temperature.
Even at the high end of those values, rapid diffusion times require that no part of the dust grain be more than a few mm thick
for Si to be able to diffuse out within a year's time.  While we expect relatively large dust grains (initial radius $a \sim 6$\op cm), the
grains are also expected to be $\sim 97\%$ porous, so this assumption is reasonable.

A further complication in estimating the $50\%$ condensation temperatures in our model is that at temperatures where the material
is not liquid, the grain sizes estimated are not insignificant ($\sim$\tp\op cm) so molecular diffusion is
unlikely to apply on the cooling timescales.  While the condensed elements are unlikely
to equilibrate chemically through the dust grains during this process, condensation timescales are still short: the collision time between
the evaporated elements and the dust grains can stretch to days or even months, but not years.
This means that it is very possible that the result of our process is
a set of veneers deposited as the gas cools, although
such a spatial segregation of elements would clearly have been erased in the rocky planets.
Further, this implies that equilibrium calculations for a solar
composition gas do not strictly apply since much of the material is already locked up in condensates.  Such
fractional condensation demonstrably occured, \emph{e.g.} CAIs.  Even though \cite{Lodders03} does not consider the
pressure and condensate dependencies, we still use her values because they
are the best estimates currently available for element condensation temperatures in the Solar nebula.

\section{Model Equations and results}
\label{sec_actual}


\subsection{Settling of compacted dust}

Even though the contracted dust grains are less sensitive to the gas motion, they remain well coupled to the gas, so there will be little slippage between the dust and gas during the rapid heating of the gas
disk and its corresponding inflation (the gas scale height $H$ increases with temperature).  The corresponding
increase in $\St_0$ for the particles
lowers their mixing height $h_m$, causing them to begin to settle even if the turbulent strength does not change substantially. 
We rewrite \Eq{terminal} as
\EQ
\dot{z}=-\frac{z}{t_{s0}} \exp \left(\frac{z^2}{2H^2}\right)=v_s. \label{falling0}
\EN
However, \Eq{falling0} is not the entire story: as the temperature drops, the gas scale height drops and the gas
disk also contracts, while \Eq{falling0} only describes the relative dust-gas motion due to the settling.

More completely, 
the equation for the time dependence of the vertical position of a settling grain is given by
\EQ
\frac{\partial}{\partial t} \left(\frac zH\right)=\frac 1H \frac{\partial z}{\partial t}-\frac{z}{H^2}\frac{\partial H}{\partial t}=
\frac{v_s}{H}. \label{falling}
\EN
\Eq{falling} implies, conveniently, that the contraction of the gas disk can be neglected as long as we scale
all lengths to $H$ and consider only settling.  If we do not consider the contraction of the gas disk,
we can discuss vertical turbulent dust mixing, or use
 \Eq{hm},  only as long as the turbulent mixing time is much shorter than the cooling time $t_c$.  This is a reasonable assumption for
our purposes: before the FUor event, the cooling time is effectively infinite; during the expansion of the gas due to the FUor heating
the dust is well coupled to the gas because its stopping time $\tau$ is much shorter than the rise time of the temperature (of
order a year); and during the event we only appeal to \Eq{hm} when the turbulent diffusion time is comparable to
the settling time, which is shorter than the cooling time.

Note also that \Eq{tau} is independent of temperature: the product $\rho_g(0) c_s$  is proportional to
$\Sigma_g$, the gas surface density, which, at least outside the FUor engine, varies slowly in time compared to the FUor or dust settling
timescales.
This means that increases in the stopping time $\tau$ due to the reduction of the sound speed as the cooling progresses are exactly cancelled by
increases in the midplane gas density due to thermal contraction of the gas disk.

We will use the subscript $-$ to refer to the dust state before the start of the FUor, and the subscript $+$ to refer to the state once
the FUor has begun.
We can solve \Eq{falling} to find that the time it takes for the top of the dust column to have fallen from its initial height $h_{m-}/H$ to $z/H<h_{m-}/H$ is
\EQ
t(z)=\frac{t_{s0+}}{2} \left[\text{E}_1 \left(\frac{z^2}{2H^2}\right)-\text{E}_1 \left(\frac{h_{m-}^2}{2H^2}\right)\right], \label{dustsettling}
\EN
where we require $z>h_{m+}$, the value of $h_m$ after the event (note that $h_{m+}$ depends on $\alpha$ and that
H is the appropriate time-varying lengthscale normalization as described above), and
\EQ
\text{E}_1(x)=-\text{Ei}(-x),
\EN
where $\text{Ei}$ is the exponential integral.

Combining \Eq{gasfraction} with \Eq{dustsettling} we can see that if the event at $t=0$ is associated with evaporating an
element from
the dust grains that does not recondense by time $t$, and does not meaningfully mix on the dust settling timescale, then the fraction of
the evaporated mass in regions which still contain dust grains at time t is well approximated by
\EQ
\widetilde{f}(t)=\frac{\erf[z(t)/2^{1/2}H]}{\erf[h_{m-}/2^{1/2}H]}, \label{frac}
\EN
see \Tab{facesof}.  We can invert \Eq{frac} to find
\EQ
z(\widetilde{f})=2^{1/2} H \erf^{-1}\left[\erf(h_{m-}/2^{1/2}H) \widetilde{f} \right], \label{z(f)}
\EN
the height to which the dust grains must settle for recondensation of volatiles to match an observed relative abundance $\widetilde{f}$.

\subsection{Simultaneous  cooling and settling}
\label{model}

We can combine Equations \ref{temp}, \ref{dustsettling}, and \ref{z(f)}  to find
\EQ
\begin{aligned}
T&\left(t\left(z\left(\widetilde{f} \tp \tp \right)\right)\right)=T_0+T_H \exp\left[-\frac{t_{s0+}}{2t_c} \times \right.\\
&\qquad \left. \left(\left[\text{E}_1\left(\frac{z^2(\widetilde{f})}{2H^2}\right)
-\text{E}_1\left(\frac{h_{m-}^2}{2H^2}\right)\right]\right)\right].
\end{aligned} \label{ugly}
\EN
This means that a given relative abundance $\widetilde{f}$ will be associated with a condensation temperature $T(\widetilde{f})$, and
by fitting \Eq{ugly} to \Eq{Earth} we can determine the dust, turbulence and FUor outburst
parameters $h_{m-}$, $t_{s0+}/t_c$, and  $T_H$.

\begin{figure}[t!]\begin{center}
\includegraphics[width=\columnwidth]{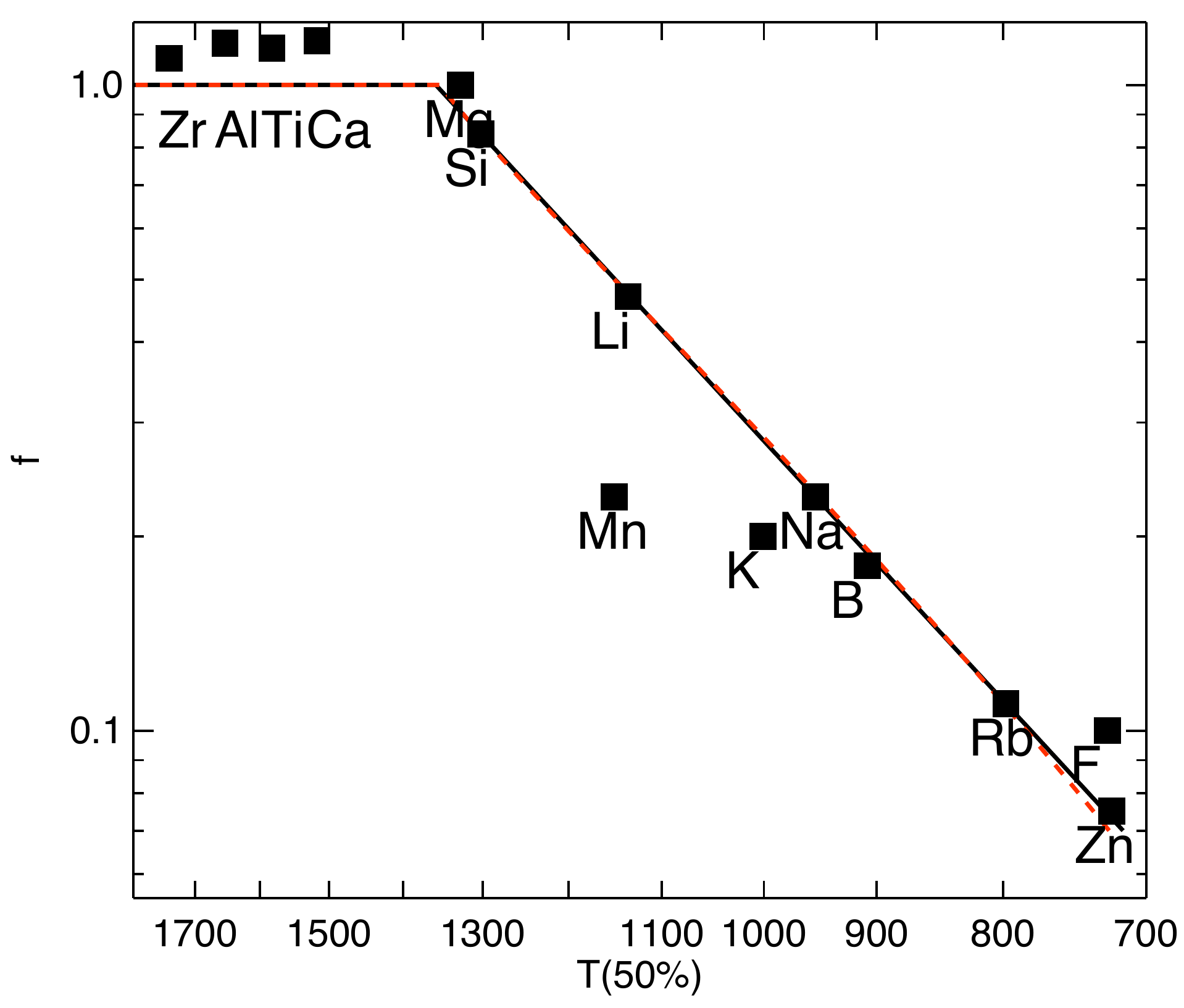}
\end{center}\caption{
Black/solid: $\ov{f}(T)$ from \Eq{Earth}.  Red/dashed: Best fit to $\ov{f}(T)$ from \Eq{ugly}. It has $h_{m-}/H=1.45$, $t_{s0+}/t_c=0.35$.
$T_0=280$\op K and $T_H=1078$\op K are imposed.
\label{Comp} }
\end{figure}

\begin{figure}[t!]\begin{center}
\includegraphics[width=\columnwidth]{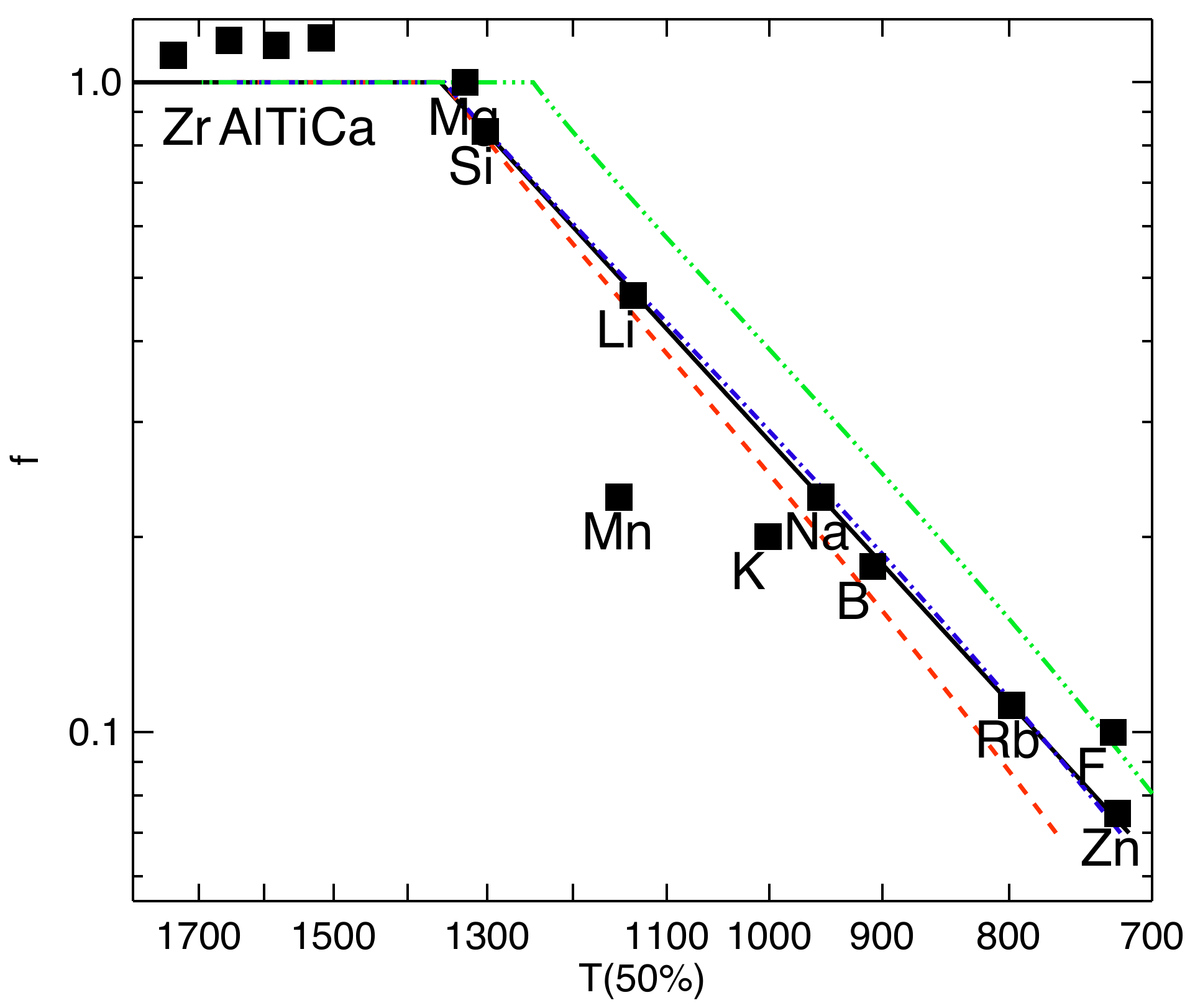}
\end{center}\caption{
Effects of varying the fit parameters.
Black/solid: $\ov{f}(T)$ from \Eq{Earth}.  Red/dashed: \Eq{ugly} using best fit values except for a $10\%$ reduction
in $t_{s0+}/tc$.  Blue/dash-dotted: \Eq{ugly} using best fit values except for a $10\%$ reduction in $\erf(h_{m-}/2^{1/2} H)$.
Green/dash-double-dotted: \Eq{ugly} using best fit values except for a $10\%$ reduction in $T_H$.}
\label{Comp2}
\end{figure}

In \Fig{Comp} we show both the assumed fit to the experimental evidence (Equation \ref{Earth}) which we attempt to match,
and our best fit model assuming
$T_0=280$\op K and $T_H=1078$\op K using \Eq{ugly}.  In \Fig{Comp2} we show the effect of varying the fit parameters.
The settling-to-cooling time ratio $t_{s0+}/t_c$ controls the slope while the heating amount $T_H$ controls the 
horizontal offset.
The initial height plays only a weak role in the depletion, although it does play a role in the initial dust porosity
and collisional velocities, see \Sec{dustparameters}.  The sensitivity of $\widetilde{f}$ on the
ratio of the settling time to the cooling time $t_{s0+}/t_c$, combined
with the lack of sensitivity of $\widetilde{f}$ on the initial dust column height $h_{m-}$ means
that the post FUor dust parameters are much better constrained than the initial dust parameters.

\begin{figure}[t!]\begin{center}
\includegraphics[width=\columnwidth]{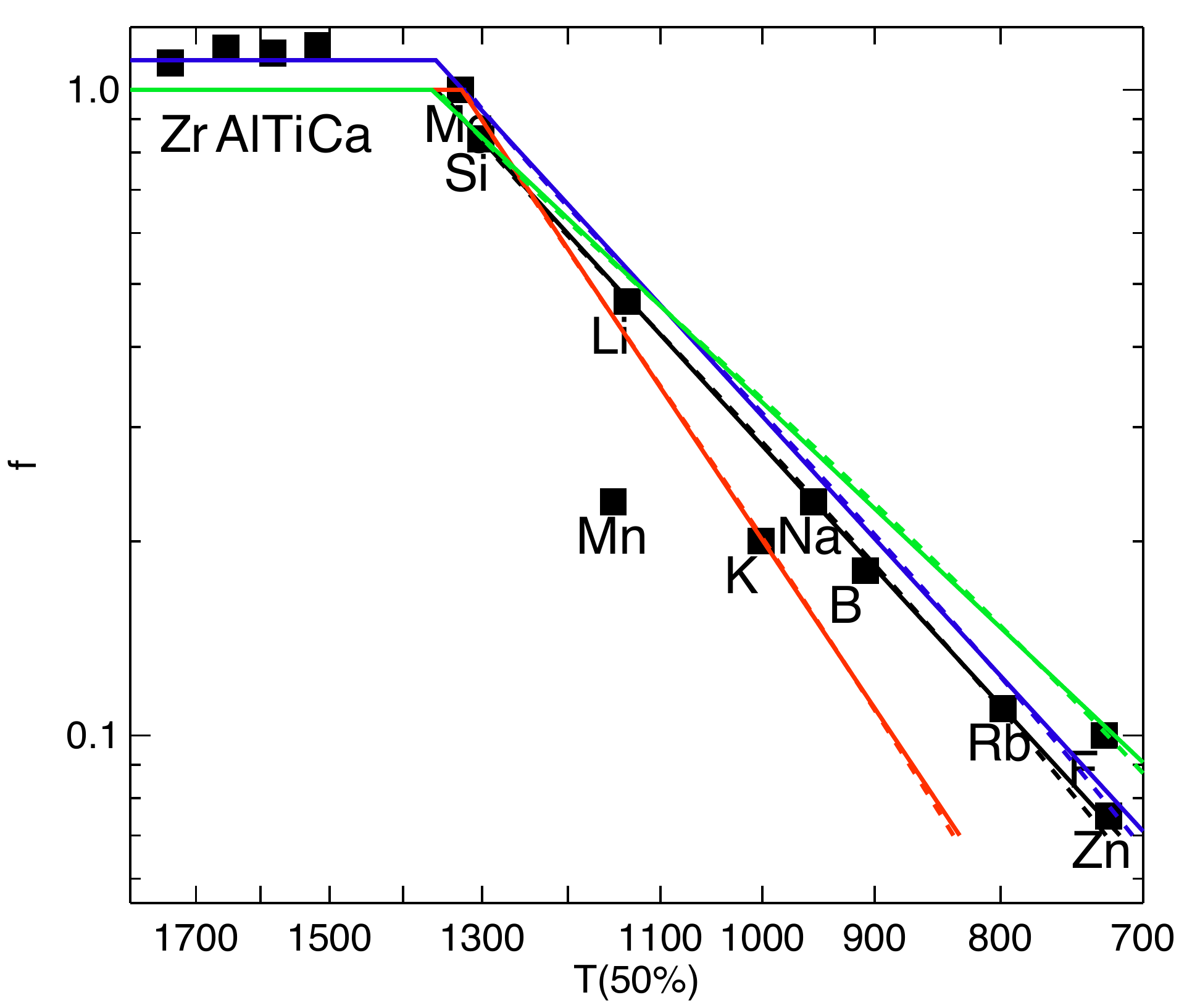}
\end{center}\caption{
Alternate fits to the data, where solid lines are
the different fits to the data and dashed lines the best match model, respectively.
Black: base fit.  Red: steep fit to the data.  Green: shallow fit to the data.
Blue: Mg is $10\%$ depleted.
\label{Comp3} }
\end{figure}

The ability
to independently change the slope and the offset allows this model
to match \Eq{Earth} for a large range of parameters.
Indeed, in \Fig{Comp3} we show the ability of the model to match
steep and shallow fits to the data (red and green), as well as matching the data assuming that the Mg is, in fact, $10\%$
depleted (blue, recalling that we equated chondritic and silicate Earth Mg abundances).
 However, the model can only fit power-law
depletions and, for example, the model cannot match an alternate fit to the depletion
data which on \Fig{Comp} goes from Si to Li, then
drops vertically to Mn, travels horizontally to Na and thence to Zn.  Finally, even though the model
lists many parameters, few of the parameters are truly unconstrained
as summarized in \Sec{final_param}.
Thus even though the model has many parameters, and is robust even when faced with the large
systematic uncertainties in
the measurements of Earth's volatile depletions, we do not feel that our model is too free.

The best fits imply dust post-densification grains with radii just under $2$\op cm, which is large enough that
molecular diffusion is unlikely to homogenize the condensed volatiles throughout the dust grains on year-length timescales.
However, it takes the disk almost exactly $1$\op yr to cool from $T=1000$\op K to $T=990$\op K.  Over one year, the dust
settles by about $4\%$: $\ov{f}(1000\tp\op \text{K}) \simeq 0.285$ and $\ov{f}(990\tp\op \rm{K}) \simeq 0.274$.  While the condensation
will create veneers on the surfaces of the grains, isotopic fractionation will not occur because isotopes condense
at nearly identical temperatures.

\subsection{Maximum turbulence strength}
\label{max_turb}

Once the grains are highly settled, turbulent mixing becomes a factor. In general, vertical Schmidt numbers are greater than 
unity \citep{Johansen06},
so the turbulent diffusion of the dust (represented by $h_{m+}$) is less important than the turbulent mixing of the
gas.  Over a settling time $t_{s0+}$, turbulence will mix columns of height $l \sim (\alpha c_s H t_{s0+})^{1/2}$.
Assuming $\erf(h_{m-}/2^{1/2}H)\simeq 0.85$, we can approximate \Eq{frac} for $z/H \ll 1$ as
\EQ
\widetilde{f}(z) \simeq 0.9~z/H.
\EN
Inserting the mixing length in the above equation we find that the abundance factor $f_m$ for which turbulent mixing becomes important
is
\EQ
f_m(\alpha, \St) = 0.9 \sqrt{\frac{\alpha_2}{\St_{0+}}}=2.2 \sqrt{\alpha_2\frac{t_{s0+}}{t_c}} \sqrt{\frac{t_c}{\text{Orb}}},
\label{fm}
\EN
where we have assumed that turbulent mixing will only becomes relevant noticeably after the temperature has
dropped below $1000$\op K and the midplane turbulence mostly decays away.
Inverting the above, we can place an upper limit on the post-MRI turbulent strength $\alpha_2$ for the model
to explain a given abundance factor $f_m$:
\EQ
\alpha_2< \left(\frac{t_c}{\text{Orb}}\right)^{-1} \left(\frac{t_c}{t_{s0+}}\right) \left(\frac{f_m}{2.2}\right)^2.
\EN
For $f_m=0.1$ and $t_c=100$\op Orb, we find $\alpha_2<6 \times 10^{-5}$, while for $f_m=0.2$ and
$t_c=50$\op Orb, we find $\alpha_2<5\times10^{-4}$.  These $f_m$ values place the top of the dust
layer well within the dead-zone.

Taking our best-fit model, $T=1000$ is met for $f \sim 0.3$.  Accordingly, the limits on $\alpha_1$ are 
approximately ten times
those for $\alpha_2$.
These are relatively stringent limits, suggesting that a plateau of $f > 0.3$ will
occur between $T\gtrsim 1000$\op K and $T=1000$\op K, with the decline in $f$ restarting for $T<1000$\op K.

\subsection{Dust parameters}
\label{dustparameters}

\subsubsection{Porosity}

In this model, the post-event grains have negligible porosity.  We can use that fact to constrain the 
porosity of the pre-event dust, as long as the turbulence strength does not vary too dramatically
because of the FUor (although it will change after the disk cools too much
to support the MRI).  When $T=1000$\op K we predict an abundance factor of $\widetilde{f}=0.29$. Using \Eq{frac}
we find that abundance factor to be associated with a height $z/H \simeq 0.31$.  Using \Eq{hm} we find that
\begin{align}
&\frac{\alpha_0}{\text{Sch}\op \St_{0-}} \simeq 1.9, \label{dustparm-}\\
&\frac{\alpha_1}{\text{Sch}\op \St_{0+}} \lesssim 0.049. \label{dustparm+}
\end{align}
Combining the above with \Eq{tau}, assuming that $\text{Sch}_-=\text{Sch}_+$, we find
\EQ
\frac{\rho_{d,s-}}{\rho_{d,s+}} \simeq  \left(\frac{0.049}{1.9}\right)^{3/2} \left(\frac{\alpha_0}{\alpha_1}\right)^{3/2}
\simeq \frac{1}{239} \times \left(\frac{\alpha_0}{\alpha_1}\right)^{3/2}. \label{poros}
\EN
As noted above, we find that the grains are settled to $z/H=0.31$ when $T=1000$\op K.  However, even if the grains
only settle to $z/H=0.5$, the density ratio is still $54^{-1} (\alpha_0/\alpha_1)^{3/2}$.  This model therefore predicts that
the initial grains have high porosity, which is associated with growth resulting from collisions between
similarly sized partners \citep{Wada08a}.

\subsubsection{Grain sizes}
\label{grainsize}

We can rewrite \Eq{tau} as
\EQ
\St_0 \simeq 3.7 \times 10^{-4} \!\left(\frac{\rho_{d,s}}{3\frac{\text{g}}{\text{cm}^3}}\right) \!\!
 \left(\frac{\Sigma_g}{1700\frac{\text{g}}{\text{cm}^{2}}}\right)^{\!\!-1}\!\! \left(\frac{a}{1\:\! \text{mm}}\right)\!, \label{Stdisk}
 \EN
 where we have assumed $\gamma=1.42$.  Combining \Eqs{dustparm+}{Stdisk} we find that the radius of the post-heating,
 non-porous grains is
 \EQ
 a_+ \sim  18 \left(\frac{\alpha_{1}}{10^{-3}}\right) \text{mm}, \label{size-falling}
 \EN
 where we have assumed that $\text{Sch} \simeq 3$, $\rho_{d,s}=3\tp \text{g\op cm}^{-3}$ and $\Sigma=1700\tp \text{g\op cm}^{-2}$
 \citep{Johansen06,Britt03,Hayashi85}.  \Eq{poros} in turn implies that if $\alpha_0 \simeq \alpha_1$, then the fluffy dust
 had size
  $a_{-} \simeq 6\op a_{+} \simeq 10\tp \text{cm}$. This
value, when applied at the top of the dust layer, is right on the edge of the transition between Epstein and Stokes drag,
and our use of the Epstein drag regime is reasonable.
Combining \Eq{size-falling} with the best-fit value $t_{s0+}/t_c=0.35$ we also find that the result is consistent with
\EQ
\left(\frac{t_c}{\text{Orb}}\right) \simeq 68 \left(\frac{\alpha_1}{10^{-3}}\right)^{-1},
\EN
a reasonable value for an FUor event for $R=1\op \text{AU}$ with $1\op \text{Orb}=1\op\text{yr}$.

Further, while dust grain collisional velocities are difficult to determine precisely \citep{Hubbard13}, the characteristic
scale for the velocities can easily be estimated \citep{OC07} as
\EQ
v_c \sim \sqrt{\alpha \op  \St} \op c_s , \label{colvel}
\EN
which when combined with \Eq{dustparm-} becomes
\EQ
v_c \sim 42 \left(\frac{\alpha_0}{10^{-3}}\right) \left(\frac{\text{Sch}}{3}\right)^{-1/2}\!\!\! \text{cm}\tp\text{s}^{-1} \label{colvel1}
\EN
for dust grains prior to the FUor (at temperatures of $280$\op K). If $\alpha_0=4\times 10^{-3}$,
then \Eq{colvel} predicts $v_c \simeq 1.6 \tp \text{m}\op \text{s}^{-1}$,
approximately half the destructive collisional velocity predicted for high porosity silicates
\citep{Wada08}, a reasonable value for the maximum characteristic dust size to be collisionally determined (dust collisions
occur over a range of velocities).

A further complication exists in that highly porous dust grains will experience compaction if they collide fast enough.
While the critical velocity for compaction, like that of fragmentation, is uncertain, it will be less than
the fragmentation speed, albeit not necessarily by that much \citep{Seizinger12}.  All the relevant velocities have
been predicted to be reasonably similar to $1$\op m/s, and depend on the chemical composition and shapes
of the constituent grains, which in the actual Solar Nebula were not perfect silicate spheres.  We use the fragmentation
speed from \citet{Wada08} because it applies to highly porous grains.

\subsection{Final parameters}
\label{final_param}

In \Tab{fit_parms} we show the best fit parameters values combining the constraints from matching the volatile depletion
 the maximum turbulent strength and the collisional velocities.
 In particular, the volatile depletion pattern constrains the heating $T_H$, the initial dust height $h_{m-}$ and the settling dynamics,
 i.e.~$t_{s0+}/t_c$ and $\alpha_2$.  Combining $h_{m-}$ with collision constraints determines $\alpha_0$.  Our assumption that
 the strong turbulence turns off at $T \simeq 1000$\op K, combined with the observed depletion, constrains how settled
 the particles are at that point, and hence $\alpha_1$, $a_{+}$ and $t_c$.  The initial size of the dust (and also
 the pre/post density ratios) are determined by the difference between $\alpha_0$ and $\alpha_1$.
 
One should recall that the initial dust column height $h_{m-}$ is very poorly constrained, and the ratio of the
initial to final grain solid density
and the initial grain size both depend sensitively on $h_{m-}$.  Indeed, an almost arbitrarily large initial porosity can
be matched by increasing $h_{m-}$.
Also, the degree of densification enters into the model through $\rho_{d,s-}/\rho_{d,s+}$.  While we have
assumed that thermal processing at $R=1$\op AU leads to complete compaction, the outcome of very long timescale
sintering has not been experimentally established.  If we instead assume incomplete compaction, then the initial dust
would have been more porous than our estimate.  Additionally, to match the settling time constraint $t_{s0+}/t_c$, the
dust grains would have had larger initial and final radii $a_-$ and $a_+$.

Our best fit cooling time $t_c = 68$\op Orb, if calculated at $R=1$\op AU,
is shorter than the estimated value for FU Orionis of about a century, but is easily close enough to be
plausible for FUors in general: we have not yet observed an FUor from onset to quiescence.  The $\alpha_0$ and $\alpha_1$
 values are also well within the standard range for MRI-active zones \citep{Blackman08}, and $\alpha_2$ is easily plausible
 for a dead-zone, especially given the non-stirring nature of dead-zone fluid motions seen in \citet{Oishi09}.  Accordingly, our
 model is in reasonable agreement with all of the (both observational and theoretical) volatile depletion, dust collision,
 FUor and MRI constraints.

\begin{table}[htbp]
\caption{Best fit parameters. \label{fit_parms}}
\centerline{\begin{tabular}{ll}
\hline
$T_0$      & $280\tp \text{K}$ \\
$T_H$     & $1078\tp\text{K}$ \\
$h_{m-}$ & $1.45 H$ \\
$t_{s0+}/t_c$ & $0.35$ \\
$t_c$       & $68\tp\text{Orb}$ \\
$\alpha_0$ & $4 \times 10^{-3}$ \\
$\alpha_1$ & $10^{-3}$ \\
$\alpha_2$ & $<9 \times 10^{-5}$ \\
$\rho_{d,s-}/\rho_{d,s+}$ & $1/30$ \\
$a_{-}$ & $6\tp\text{cm}$ \\
$a_{+}$ & $18 \tp\text{mm}$ \\
\hline
\end{tabular}}
\end{table}

\section{Extension to Mars, Chondrites and beyond}
\label{extension}

Our model applies as long as the FUor (or other heating event) is hot enough to rapidly melt or sinter the dust grains.
In what follows we will assume that $T \sim R^{-1/2}$, appropriate for
an irradiated disk, and that the surface density scales as $\Sigma \sim R^{-1}$.

\subsection{Mars}
\label{sec:mars}

Mars is at $R \simeq 1.5$\op AU,  which implies that the peak temperature reached at Mars' location was $T \simeq 1100$\op K.
Extrapolating the results from \cite{Poppe03}, measurable sintering for SiO$_2$ grains is expected after a decade:
the neck between $0.7 \mu m$ grains would grow to about $0.3 \mu m$.  Note however
that we have extrapolated well below the temperatures and well above the timescales tested in that work, and that
the result is very sensitive to the excitation energy.
Clearly densification through various processes is possible, but whether it will be adequate
to significantly change the aerodynamical properties of the dust will depend on the chemical makeup
of the grains.  Certainly, however, an FUor is unlikely to result in completely solid silicate grains at Mars' position (or beyond),
and there is no particular reason to assume that Mars formed simultaneously with the Earth.  However, we can
explore the consequences of our model at Mar's location and
check whether or not observations of Martian potassium to thorium ratios are consistent with Mars having formed during
an FUor event.

We can apply \Eq{colvel} while assuming that
the dust grain size is set collisionally by a constant peak collisional velocity.  If we also set a turbulence strength $\alpha$
that is independent of radial position, we find that
\EQ
\St_{\text{Mars}} \simeq 1.2 \tp \St_{\text{Earth}},
\EN
which when combined with \Eq{dustparm-} implies a value of  $h_{m-}=1.38$, similar to that for dust
at Earth's position.  Note that $\St_{\text{Mars}}$ and $\St_{\text{Earth}}$ are normalized to different Keplerian
frequencies, so the stopping time ratio is $\tau_{\text{Mars}} \simeq 3\op \tau_{\text{Earth}}$, and the grains at Mars' position have
 double the value of $(\rho_{d,s}\op a)$.

\begin{figure}[t!]\begin{center}
\includegraphics[width=\columnwidth]{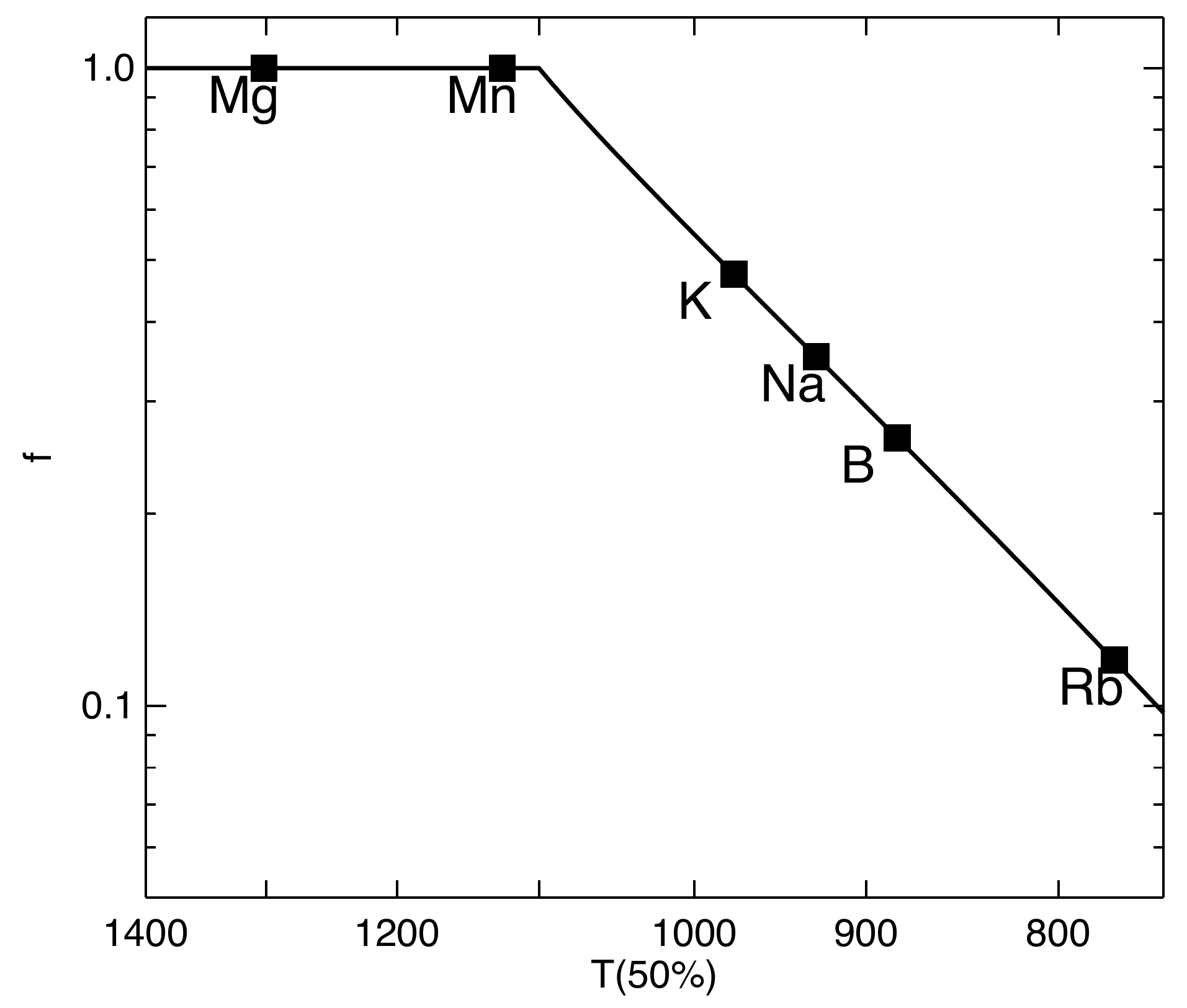}
\end{center}\caption{
Predicted volatile depletion for Mars.  Condensation temperatures are decreased by $25$\op K from \cite{Lodders03}
due to the decrease in ambient pressure.
\label{mars} }
\end{figure}

Using K/Th ratios from \cite{Taylor06}, we can estimate that the normalized potassium abundance for Mars is $1.7$ that of Earth.
Assuming that significant densification does occur, and
noting that the pressures experienced by dust grains at Mars' orbit are approximately one third those
at Earth's orbit, which we approximate by decreasing condensation temperatures by $25$\op K, as discussed in \Sec{Earth'sCompostion},
we can apply our model to fit Mars' potassium depletion.
Our model matches
\EQ
\widetilde{f}_{\text{Mars}} (981\op \text{K})=1.7\widetilde{f}_{\text{Earth}}(1006\op \text{K})
\EN
for $t_{s0+}/t_c=0.23$ and an initial/final dust grain density ratio of $\sim 1/4$.   See \Fig{mars} for the predicted depletions.
If we assume that the initial porosity
of the grains that became Mars was the same as those that became Earth, the contracted state of the pre-Mars
grains was about $87\%$ porous.  Further, the linear contraction during the densification phase was $(1/4)^{1/3} \simeq 
0.62$, which can be used to constrain the densification process.
This degree of densification is plausible, so Mars is consistent with having formed in the same FUor as Earth.

The Martian abundances are, however, poorly constrained, and similar to the Earth's \citep{Peplowski11}.
Our model cannot fit potassium abundances less than
$1.4$ times those of the Earth.  At that value, and assuming that the pre-Mars grains have the same initial porosity as the pre-Earth
grains, there needs to have been complete densification of the pre-Mars grains.

\subsection{Chondrites}

FUor events and our model have some consequences for chondrule and chondrite formation.  Most simply,
FUors will raise temperatures up to $650$\op K at the top of the dust layer (and possibly higher at the midplane) 
out to about $3.7$\op AU, enough to destroy even the high-$T$ presolar component \citep{Mendybaev02,Huss03}, a constraint
that models of chondrite formation have to consider.

At $2.2$\op AU, near the inner edge of the asteroid belt,
the peak temperature reached would be nearer to $850$\op K, still too low to melt fluffy grains,
but potentially high enough for the short circuit instability to trigger \citep{Hubbard12,McNally13}.
The short circuit instability causes very localized
heating events in current sheets which can reach temperatures appropriate for chondrule formation, and certainly would
result in rapid Stokes number changes by melting and reforming the grains.

\subsection{Ices and dust destruction}

Similar processes could be expected to also occur significantly beyond the ice line, as FUors would not only
change the porosity of silicate grains, but also that of ice grains.  Near the ice line, ice grains would be entirely destroyed
during an outburst, but far enough out, beyond $10$\op AU or about Saturn's orbit, the peak temperature reached during an FUor would be near
the temperature associated with water ice melting, and an analogue to our mechanism would reduce the dust grain
porosity.  This would lead to enhanced dust settling and potentially the streaming instability and rapid planetesimal formation.
Even in intermediate regions, near Jupiter's orbit, the evaporation of ices, followed by recondensation onto the silicate dust seeds,
would lead to more compact grains than would result from collisional agglomeration.  Similar
evaporation and recondensation processes have been suggested as routes towards planet formation \citep{Ros13}, and could,
in fact, also lead to planet formation inside the orbital position where silicate grains are destroyed during an FUor.  That would
occur inside of $0.7$\op AU, nearly exactly Venus' orbit, suggesting that the destruction of silicate grains during outbursts
 could have been important
for the formation of both Venus and Mercury.

\section{Conclusions}
\label{conclusions}

We have presented a model for the depletion of Earth's volatiles as measured with respect to Solar abundances.
The model relies on the spatial segregation of volatiles in the gas phase with dust grains through the vertical settling
of the dust after an FU Orionis type outburst heats the disk adequately to evaporate the volatiles.  The heating causes
highly porous grains to contract and settle, abandoning much of the volatiles in the upper reaches of the disk, but only after
isotopic reequilibration with those volatiles.
Finally, the Streaming Instability gathers the settled grains, triggering gravitational instability and leading to direct collapse
to planetesimals.
Our model explains not only the Earth's volatile depletion, but also the trickier lack of an isotopic signature in the potassium depletions
because it provides long enough timescales for the different isotopes to equilibrate independently.

This model predicts
that collisional agglomeration naturally creates high porosity fluffy dust grains, and that the planetesimals that
became Earth formed early 
and in situ.
The model predicts volatile depletions for all planets formed within the critical annulus inside of which
the background disk was cool enough to allow the formation of fluffy dust grains, but
FU Orionis type events were hot enough to melt or sinter the dust grains, causing them to become compact
and to settle.  In the case of the proto-Solar nebula, this extends from about Venus' orbit to just about Mars', although
the implications for Mars are weaker because the heating is unlikely to completely contract the dust grains.

Our models ties together a large number of processes which still have large systematic uncertainties.
We have observed only a handful of FUors, none to completion, and the growth of
dust grains of unknown geometry and surface chemistry has been modeled with only gross simplifications.
Indeed, it is only recently that the question of icy surfaces has been treated in some detail alongside SiO$_2$
grains.  Accordingly, we have described the simplest version of our model to explain it qualitatively, and to show
that it holds quantitative promise.  Some extensions are unlikely to pose significant problems: planets must aggregate
materials from a significant annulus, rather than merely a single position.  This will add together multiple depletion patterns,
but if a model predicts a depletion pattern that matches the Earth's at one radial position, feeding from an annulus of modest width near
that position should not change the final result dramatically.

Like all studies of planet formation, our model would be significantly improved by a better understanding of the outcome
of dust-dust collisions.  Along with studies of chondrites, we would additionally benefit from systematic study
of the diffusion of volatile elements through silicates under hot nebular conditions.  Finally, our model makes the study of long
(month or year) timescale sintering of Solar Nebula minerals important. 
The experiences of \citet{Poppe03} shows that such studies need to be performed
in zero-G because gravity acts to compress materials.

We have also assumed that $R=1$\op AU lies outside of the FUor engine, which allows us to neglect
the mass flow onto the star that powers the FUor, keeping our material from moving radially on the timescales
we consider.  The extent of FUor engines
is not known (observations have not yet resolved them), but they may well extend beyond $R=1$\op AU \citep{Zhu10}.  While our model
can adjust for this by moving radially outwards, this will inevitably result in lower temperatures unless the FUor is brighter.
However, the luminosity jump we associate with the FUor is on the strong end
of observations, so a large outwards change in radial positioning is probably not possible.  Finally, planet migration raises strong
questions about the link between the initial and final position of a planet which are beyond the scope of this work.
Interestingly though, the Earth's depletion
pattern suggests that it must have been made reasonably near $1$\op AU.  Too much farther out, and even an FUor
wouldn't be able to raise the temperature enough to evaporate volatiles, while too close in, and even quiescent disks will be too hot.

The implication that collisionally grown dust in protoplanetary disks has high porosity means that thermal
reduction of dust grain porosity will be a significant player in planet formation.  By growing dust grains
in a fractal, high porosity manner, even high mass dust grains remain tightly coupled to the gas, experiencing
relatively low collisional velocities: low porosity dust grains of the same mass would experience destructive collisions.
Subsequent heating and grain contraction then led to highly settled
dust grains that are subject to the streaming instability, leading to direct gravitational collapse of the dust disk
to form planetesimals.  While we appeal to FU Orionis type events in this model, other sources of heating could play a role. 

It is difficult to make specific predictions for our model in the Solar System because it is likely that many objects formed
during quiescent phases.  The large number of meteorite classes suggests that some probably did form close in time
to an accretion event, so preserving pre-solar grains during FUors is a problem, albeit one which applies independently of our model.  Further, while we believe FUor events likely promote
volatile-depleted rocky planet formation in  extra-solar, often heavily embedded, systems, observationally
testing that hypothesis is not currently possible in part because of the rarity of FUors.  Nonetheless, our model states that an initial supply of
highly porous dust grains will be compactified.  This should have observational implications, particularly when applied to icy
grains on large enough orbits that their their effect on disk spectra occur in annuli which can be resolved.  This will require
an adequate survey of protostellar systems that future FUors can be compared with their previous, quiescent state.

\acknowledgements

We thank Mordecai-Mark Mac Low and especially the two referees for their detailed critiques that went far beyond
the minimum effort and helped immensely in improving not only the manuscript's legibility, but also the science.
This research has made use of the National Aeronautics and Space AdministrationÕs Astrophysics Data System Bibliographic Services.
The work was supported by National Science Foundation, Cyberenabled Discovery Initiative grant AST08-35734, 
National Aeronautics and Space Administration grant NNX10AI42G (DSE), 
and a Kalbfleisch Fellowship from the American Museum of Natural History.


\begin{thebibliography}

\bibitem[Albar{\`e}de(2009)]{Albarede09} Albar{\`e}de, F.\ 2009, 
\nat, 461, 1227 

\bibitem[Anders(1964)]{Anders64} Anders, E.\ 1964, \ssr, 3, 583 

\bibitem[Armitage et al.(2001)]{Armitage01} Armitage, P.~J., 
Livio, M., \& Pringle, J.~E.\ 2001, \mnras, 324, 705 

\bibitem[Bae et al.(2013)]{Bae13} Bae, J., Hartmann, L., Zhu, 
Z., \& Gammie, C.\ 2013, \apj, 764, 141 

\bibitem[Balbus 
\& Hawley(1991)]{BH91} Balbus, S.~A., \& Hawley, J.~F.\ 1991, \apj, 376, 214 

\bibitem[Birnstiel et 
al.(2012)]{Birnstiel12} Birnstiel, T., Klahr, H., \& Ercolano, B.\ 2012, \aap, 539, A148

\bibitem[Blackman et al.(2008)]{Blackman08} Blackman, E.~G., 
Penna, R.~F., \& Varni{\`e}re, P.\ 2008, New Astronomy, 13, 244 

\bibitem[Britt 
\& Consolmagno(2003)]{Britt03} Britt, D.~T., \& Consolmagno, G.~J.\ 2003, Meteoritics and Planetary Science, 38, 1161 

\bibitem[Bond et al.(2010)]{Bond10} Bond, J.~C., Lauretta, 
D.~S., \& O'Brien, D.~P.\ 2010, Icarus, 205, 321 

\bibitem[Cassen(1996)]{Cassen96} Cassen, P.\ 1996, Meteoritics and Planetary Science, 31, 793 

\bibitem[Cassen(2001)]{Cassen01} Cassen, P.\ 2001, Meteoritics and Planetary Science, 36, 671 

\bibitem[Ciesla(2008)]{Ciesla08} Ciesla, F.~J.\ 2008, Meteoritics and Planetary Science, 43, 639 

\bibitem[Dominik et al.(2007)]{Dominik07} Dominik, C., Blum, J., 
Cuzzi, J.~N., \& Wurm, G.\ 2007, Protostars and Planets V, 783 

\bibitem[Dunham et al.(2008)]{Dunham08} Dunham, M.~M., Crapsi, 
A., Evans, N.~J., II, et al.\ 2008, \apjs, 179, 249 

\bibitem[Ebel 
\& Grossman(2000)]{Ebel00} Ebel, D.~S., \& Grossman, L.\ 2000, \gca, 64, 339 

\bibitem[Ebel(2006)]{Ebel06} Ebel, D.~S.\ 2006, Meteorites and 
the Early Solar System II, 253 

\bibitem[Ebel et 
al.(2008)]{Ebel08} Ebel, D.~S., Weisberg, M.~K., Hertz, J., \& Campbell, A.~J.\ 2008, Meteoritics and Planetary Science, 43, 1725 

\bibitem[Elser et al.(2012)]{Elser12} Elser, S., Meyer, M.~R., 
\& Moore, B.\ 2012, Icarus, 221, 859 

\bibitem[Evans et al.(2009)]{Evans09} Evans, N.~J., II, Dunham, 
M.~M., J{\o}rgensen, J.~K., et al.\ 2009, \apjs, 181, 321 

\bibitem[Flock et 
al.(2013)]{Flock13} Flock, M., Fromang, S., Gonz{\'a}lez, M., \& Commer{\c c}on, B.\ 2013, \aap, 560, A43 

\bibitem[Freer(1981)]{Freer81} Freer, R.\ 1981, Contributions 
to Mineralogy and Petrology, 76, 440 

\bibitem[Gammie(1996)]{Gammie96} Gammie, C.~F.\ 1996, \apj, 457, 
355 

\bibitem[Hartmann 
\& Kenyon(1996)]{Hartmann96} Hartmann, L., \& Kenyon, S.~J.\ 1996, \araa, 34, 207 

\bibitem[Hayashi et al.(1985)]{Hayashi85} Hayashi, C., Nakazawa, 
K., \& Nakagawa, Y.\ 1985, Protostars and Planets II, 1100 

\bibitem[Hubbard et al.(2012)]{Hubbard12} Hubbard, A., McNally, 
C.~P., \& Mac Low, M.-M.\ 2012, \apj, 761, 58

\bibitem[Hubbard(2013)]{Hubbard13} Hubbard, A.\ 2013, 
arXiv:1303.6639 

\bibitem[Huss et al.(2003)]{Huss03} Huss, G.~R., Meshik, 
A.~P., Smith, J.~B., \& Hohenberg, C.~M.\ 2003, \gca, 67, 4823 

\bibitem[Humayun 
\& Clayton(1995)]{Humayun95} Humayun, M., \& Clayton, R.~N.\ 1995, \gca, 59, 2131

\bibitem[Johansen et al.(2006)]{Johansen06} Johansen, A., Klahr, 
H., \& Mee, A.~J.\ 2006, \mnras, 370, L71 

\bibitem[Johansen et al.(2007)]{Johansen07} Johansen, A., Oishi, 
J.~S., Mac Low, M.-M., et al.\ 2007, \nat, 448, 1022

\bibitem[Krauss 
\& Wurm(2005)]{Krauss05} Krauss, O., \& Wurm, G.\ 2005, \apj, 630, 1088 

\bibitem[Lambrechts 
\& Johansen(2012)]{Lambrechts12} Lambrechts, M., \& Johansen, A.\ 2012, \aap, 544, A32 

\bibitem[LaTourrette 
\& Wasserburg(1998)]{Latourette98} LaTourrette, T., \& Wasserburg, G.~J.\ 1998, Earth and Planetary Science Letters, 158, 91 

\bibitem[Lodders(2003)]{Lodders03} Lodders, K.\ 2003, \apj, 591, 
1220 

\bibitem[Maxey(1987)]{Maxey87} Maxey, M.~R.\ 1987, Journal of 
Fluid Mechanics, 174, 441 

\bibitem[McDonough(2003)]{McDonough03} McDonough, W.~F.\ 2003, 
Treatise on Geochemistry, 2, 547 

\bibitem[McNally et al.(2013)]{McNally13} McNally, C.~P., 
Hubbard, A., Mac Low, M.-M., Ebel, D.~S., 
\& D'Alessio, P.\ 2013, \apjl, 767, L2 

\bibitem[Mel{\'e}ndez et al.(2009)]{Mel09} Mel{\'e}ndez, J., 
Asplund, M., Gustafsson, B., \& Yong, D.\ 2009, \apjl, 704, L66 

\bibitem[Mendybaev et al.(2002)]{Mendybaev02} Mendybaev, R.~A., 
Beckett, J.~R., Grossman, L., et al.\ 2002, \gca, 66, 661 

\bibitem[Metzler et al.(1992)]{Metzler92} Metzler, K., Bischoff, 
A., \& Stoeffler, D.\ 1992, \gca, 56, 2873 

\bibitem[Morbidelli et al.(2012)]{Morbidelli12} Morbidelli, A., 
Lunine, J.~I., O'Brien, D.~P., Raymond, S.~N., 
\& Walsh, K.~J.\ 2012, Annual Review of Earth and Planetary Sciences, 40, 251 

\bibitem[Natta et 
al.(2006)]{Natta06} Natta, A., Testi, L., \& Randich, S.\ 2006, \aap, 452, 245 

\bibitem[Oishi 
\& Mac Low(2009)]{Oishi09} Oishi, J.~S., \& Mac Low, M.-M.\ 2009, \apj, 704, 1239 

\bibitem[Ormel 
\& Cuzzi(2007)]{OC07} Ormel, C.~W., \& Cuzzi, J.~N.\ 2007, \aap, 466, 413 

\bibitem[Ormel et 
al.(2007)]{Ormel07} Ormel, C.~W., Spaans, M., \& Tielens, A.~G.~G.~M.\ 2007, \aap, 461, 215 

\bibitem[Palme(2000)]{Palme00} Palme, H.\ 2000, \ssr, 92, 237

\bibitem[Palme 
\& O'Neill(2003)]{Palme03} Palme, H., \& O'Neill, H.~S.~C.\ 2003, Treatise on Geochemistry, 2, 1 

\bibitem[Peplowski et al.(2011)]{Peplowski11} Peplowski, P.~N., 
Evans, L.~G., Hauck, S.~A., et al.\ 2011, Science, 333, 1850 

\bibitem[Poppe(2003)]{Poppe03} Poppe, T.\ 2003, Icarus, 164, 
139 

\bibitem[Richter(2004)]{Richter04} Richter, F.~M.\ 2004, \gca, 
68, 4971 

\bibitem[Ros 
\& Johansen(2013)]{Ros13} Ros, K., \& Johansen, A.\ 2013, arXiv:1302.3755 

\bibitem[Rubin(1999)]{Rubin99} Rubin, A.\ 1999, \gca, 63, 2281 

\bibitem[Seizinger et 
al.(2012)]{Seizinger12} Seizinger, A., Speith, R., \& Kley, W.\ 2012, \aap, 541, A59 

\bibitem[Shakura 
\& Sunyaev(1973)]{SS73} Shakura, N.~I., \& Sunyaev, R.~A.\ 1973, \aap, 24, 337 

\bibitem[Takeuchi 
\& Lin(2002)]{Takeuchi02} Takeuchi, T., \& Lin, D.~N.~C.\ 2002, \apj, 581, 1344 

\bibitem[Taylor et al.(2006)]{Taylor06} Taylor, G.~J., Stopar, 
J.~D., Boynton, W.~V., et al.\ 2006, Journal of Geophysical Research 
(Planets), 111, 3 

\bibitem[Wada et al.(2008a)]{Wada08a} Wada, K., Tanaka, H., 
Suyama, T., Kimura, H., 
\& Yamamoto, T.\ 2008, Lunar and Planetary Institute Science Conference Abstracts, 39, 1545 

\bibitem[Wada et al.(2008b)]{Wada08} Wada, K., Tanaka, H., 
Suyama, T., Kimura, H., \& Yamamoto, T.\ 2008, \apj, 677, 1296 

\bibitem[Wasson 
\& Chou(1974)]{Wasson74} Wasson, J.~T., \& Chou, C.-L.\ 1974, Meteoritics, 9, 69 

\bibitem[Wasson(1985)]{Wasson85} Wasson, J.~T.\ 1985, New York, 
W.~H.~Freeman and Co., 1985, 274 p.,  

\bibitem[Wasson 
\& Kallemeyn(1988)]{Wasson88} Wasson, J.~T., \& Kallemeyn, G.~W.\ 1988, Royal Society of London Philosophical Transactions Series A, 325, 535 

\bibitem[Wasson(2008)]{Wasson08} Wasson, J.~T.\ 2008, Icarus, 
195, 895 

\bibitem[Weidenschilling(1977)]{Weidenschilling77} Weidenschilling, 
S.~J.\ 1977, \mnras, 180, 57 

\bibitem[Wood et al.(2010)]{Wood10} Wood, B.~J., Halliday, 
A.~N., \& Rehk{\"a}mper, M.\ 2010, \nat, 467, 

\bibitem[Youdin 
\& Goodman(2005)]{Youdin05} Youdin, A.~N., \& Goodman, J.\ 2005, \apj, 620, 459 

\bibitem[Youdin 
\& Johansen(2007)]{Youdin07} Youdin, A., \& Johansen, A.\ 2007, \apj, 662, 613 

\bibitem[Zhu et al.(2010)]{Zhu10} Zhu, Z., Hartmann, L., 
Gammie, C.~F., et al.\ 2010, \apj, 713, 1134 


\bibitem[Zsom et 
al.(2010)]{Zsom2010} Zsom, A., Ormel, C.~W., G{\"u}ttler, C., Blum, J., \& Dullemond, C.~P.\ 2010, \aap, 513, A57 

\end{thebibliography}
\end{document}